\pdfoutput=1
\documentclass[compsoc,conference,a4paper,10pt,times]{IEEEtran}
\IEEEoverridecommandlockouts
\usepackage{cite}
\usepackage{amsmath,amssymb,amsfonts}
\usepackage{graphicx}
\usepackage{textcomp}
\usepackage{xcolor}
\usepackage{booktabs,tabularx,array,makecell,xltabular}
\usepackage[caption=false,font=footnotesize]{subfig}
\usepackage{tikz}
\usetikzlibrary{arrows.meta,positioning}
\usepackage[colorlinks=true,urlcolor=black,linkcolor=black,citecolor=black]{hyperref}
\newcolumntype{P}[1]{>{\raggedright\arraybackslash}p{#1}}
\newcolumntype{Y}{>{\raggedright\arraybackslash}X}

\newcommand{\finding}[2]{\par\smallskip\noindent\fcolorbox{black!60}{black!5}{\parbox{\dimexpr\linewidth-2\fboxsep-2\fboxrule\relax}{\small\textbf{Finding #1.} #2}}\par\smallskip}
\newcommand{\rec}[2]{\par\noindent\textbf{#1}\enspace #2\par}
\begin{document}

\title{SoK: Splitting the Trigger or Adding Attackers? Disentangling Distributed Backdoor Attacks in Federated Learning}

\author{\IEEEauthorblockN{Zhuotao Lian\thanks{Corresponding author: Zhuotao Lian.}}
\IEEEauthorblockA{Graduate School of Advanced\\Science and Engineering\\
Hiroshima University\\
Higashi-Hiroshima, Japan}
\and
\IEEEauthorblockN{Qingkui Zeng}
\IEEEauthorblockA{\\Tongling University\\
Tongling, China}
\and
\IEEEauthorblockN{Weiyu Wang}
\IEEEauthorblockA{Graduate School of\\Science and Engineering\\
Hosei University\\
Tokyo, Japan}}

\maketitle

\begin{abstract}
Distributed backdoor attacks split a trigger among several federated learning clients, and a line of work reports that this makes backdoors stronger, more persistent, and stealthier than a single attacker with the full trigger. We show that this belief rests on a confounded comparison. The distributed attack also uses more attackers, and the field has credited their extra poisoning budget to the split. We systematize nineteen primary studies and audit fifteen papers that use DBA as a baseline. None compares a split trigger with the same attackers using the full trigger, and none tests whether single pieces stay effective. We then separate the two factors with a complete-trigger baseline that keeps the attackers fixed, in 875 paired federated runs. Three insights follow. First, the advantage is a budget effect: adding attackers raised attack success in every seed, by 7.5 to 83.4 percentage points on average, while splitting among the same attackers mattered much less once both were tuned alike. Second, the layout decides whether splitting helps. DBA's compact split brings no gain, but the same pieces at the image corners beat the complete trigger on CIFAR-10, grayscale CIFAR-10, and GTSRB, and they survive Neural Cleanse better. Third, split training produces several partially working triggers rather than one composite trigger, so defenses tested only on the full trigger measure the wrong object. We turn these insights into a reporting checklist for attack and defense evaluations.
\end{abstract}


\section{Introduction}\label{sec:intro}
Federated learning (FL) trains a shared model from updates that clients compute on local data~\cite{mcmahan2017}. The aggregator sees only these updates, so a malicious client can implant a backdoor, i.e., an attacker-chosen behavior that a trigger activates, while the model keeps its normal accuracy~\cite{bagdasaryan2020,bhagoji2019lens}. A realistic adversary controls more than one client, e.g., a compromised device fleet or Sybil identities in cross-device FL~\cite{Fung2020FoolsGold}, or colluding organizations in cross-silo FL.

\begin{figure*}[t]
\centering
\includegraphics[width=\textwidth]{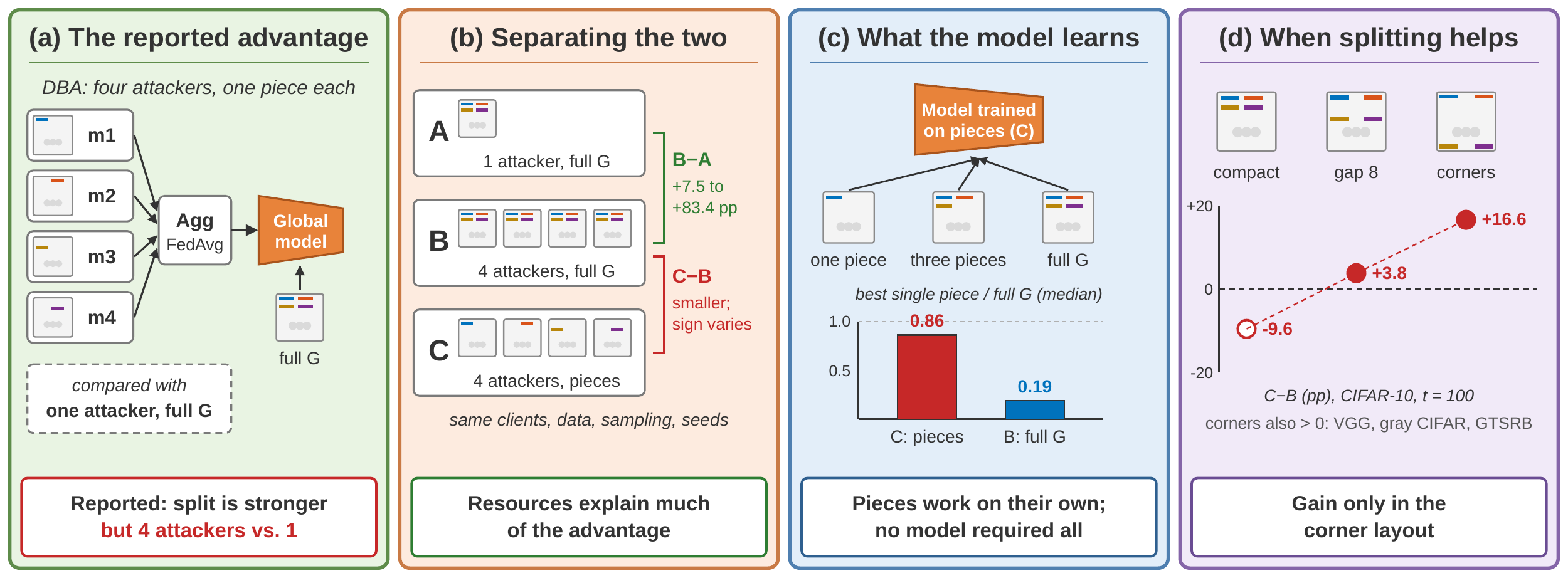}
\caption{Overview. (a) DBA reports that four attackers, each training on one piece of the trigger, beat one attacker with the full trigger $G$. This comparison changes the number of attackers and the split at once. (b) Our paired arms separate the two. B$-$A is the effect of adding attackers (means over 30 cells), and C$-$B is the effect of splitting among the same attackers. C$-$B is smaller and changes sign with the layout, the allocation, and the local training. (c) Models trained on pieces respond to a single piece at a median 0.86 of their full-trigger ASR, against 0.19 for models trained on $G$. No model required all pieces. (d) On CIFAR-10/ResNet-18, the composition effect after 100 benign rounds turns positive as the pieces move apart. The corner layout also helps on VGG-11, grayscale CIFAR-10, and GTSRB.}\label{fig:system}
\end{figure*}

The distributed backdoor attack (DBA) of Xie et al.~\cite{xie2020dba} made this setting its own attack design. Each malicious client trains on one piece of a global trigger, and the aggregated model responds to the full trigger (Fig.~\ref{fig:system}a). DBA was reported to be more effective, more persistent, and harder to detect than a single attacker with the full trigger. Many follow-up attacks build on this idea, with combinatorial pieces~\cite{liu2024fcba}, redundant shares~\cite{yang2025ssrdba}, adaptive pieces~\cite{feng2025sadba}, graph and temporal triggers~\cite{Xu2024FedGNNBackdoors,yang2024optgdba,abad2024timebandits}, decentralized topologies~\cite{liu2026dbadfl}, and coordinated updates~\cite{yang2022cooperative,lyu2025coba,alharbi2024cba}. DBA is also a standard attack in the evaluation of FL defenses.

We argue that this line of work rests on a comparison that cannot support its main claim. A distributed attack differs from the single-attacker baseline in two ways at once. It splits the trigger, and it uses more attackers, which submit more poisoned updates. The original comparison matched the trigger pixels but not the attackers~\cite{xie2020dba}, and later work kept this design. The field has therefore measured the attacker's budget and credited it to the split. Two questions follow, and neither has been answered: what does splitting itself contribute, and what does a model trained on pieces actually learn?

This SoK answers both. We first systematize the distributed backdoor literature and audit which comparisons support its claims. We then separate the two factors with paired experiments. Our key design is a \emph{complete-trigger baseline with the same attackers} (arm B in Fig.~\ref{fig:system}b). Against a single attacker (A), B measures the effect of adding attackers. Against the split trigger (C), B measures the effect of splitting among the same attackers. We ask three questions. \textbf{RQ1:} Where does the reported advantage come from? \textbf{RQ2:} What does a split-trained model learn? \textbf{RQ3:} When does splitting help?

The answers change how distributed backdoors should be understood and evaluated. The advantage comes mainly from the attackers: adding attackers raised attack success in every seed of every configuration, while splitting among the same attackers mattered much less once both were tuned. Splitting is not useless, however. Its value depends on where the pieces are placed. With DBA's compact layout it brings no gain, but the same pieces placed at the image corners beat the complete trigger on three datasets and also survive trigger reverse engineering better. Finally, split training does not produce one composite trigger. It produces several partially working triggers, so a defense that is tested only on the full trigger measures the wrong object.

\paragraph{Contributions}
\begin{itemize}
\item \textbf{Systematization} (Section~\ref{sec:sys}). We code nineteen primary studies on separate axes: what is distributed, how it reaches the model, who participates, the topology, the coordination channel, the post-attack horizon, and the tested defenses. We identify four coordination failure modes that the constructions address: coverage, detectability, retention, and interference. We also define necessary composition, independently effective pieces, post-stop persistence, and the attacker budget, which make the claims of different papers comparable.
\item \textbf{Claims audit} (Section~\ref{sec:evidence}). We map each claimed advantage to the comparison that would support it. No coded study compares a split trigger with the same attackers using the full trigger. In a sample of fifteen papers that use DBA as a baseline, none tests whether single pieces remain effective.
\item \textbf{Disentangling experiment} (Section~\ref{sec:exp}). We introduce paired arms that separate the resource, composition, and redundancy effects, and run them in 875 federated runs over nine dataset and model configurations, four aggregation defenses, three server-side repairs, three piece layouts, a DBA-style scaled setting, and an FCBA-style allocation.
\item \textbf{Insights and a reporting checklist} (Sections~\ref{sec:exp} and~\ref{sec:lessons}). We show that attacker resources drive the reported advantage, that the value of splitting depends on the layout, and that split training produces several partially working triggers. We turn these results into a minimum reporting checklist for attack and defense evaluations.
\end{itemize}

\paragraph{Key findings}
\begin{itemize}
\item \emph{The advantage is a budget effect.} Adding attackers raised attack success in every seed of all 30 cells where we measured it, by 7.5 to 83.4 points on average. Wherever both arms had the same tuning budget, splitting among the same attackers changed attack success by only $-16.5$ to $+18.5$ points, and in a DBA-style scaled setting a single attacker that injects four times came within 6.1 points of four attackers (Finding~2).
\item \emph{Split training creates several triggers, not one.} Models trained on pieces respond to their best single piece at a median 86\% of their full-trigger success, against 19\% for models trained on the full trigger (Finding~3).
\item \emph{The layout decides whether splitting helps.} With the pieces at the image corners, the split trigger beat the complete trigger in every seed on CIFAR-10 with two architectures, on grayscale CIFAR-10, and on GTSRB, by up to 18.5 points after 100 benign rounds. With DBA's compact layout, it did not (Finding~4).
\item \emph{Corner-split backdoors are harder to repair.} Aggregation defenses filtered split and complete-trigger updates alike. Neural Cleanse suppressed single pieces together with the full trigger, but after this repair the corner-split backdoor kept 23.0 points more attack success than the complete one, in every seed (Finding~5).
\end{itemize}
Our experimental findings hold for spatial patch triggers in image classification with FedAvg. Our goal is not to fault individual papers, most of which followed the evaluation practice of their time, but to make clear what the evidence supports.

\section{Background and Scope}\label{sec:background}
\paragraph{Federated training} In round $t$, the aggregator selects clients $\mathcal{S}_t$, sends them the global parameters $w_t$, and combines the returned updates, $w_{t+1}=\mathrm{Agg}(w_t,\{u_{i,t}: i\in\mathcal{S}_t\})$. In server-based FL, the aggregator is a single party. In decentralized FL, peers mix models over a communication graph. A backdoor attacker controls a set $\mathcal{M}$ of clients. It wants the global model to keep its clean accuracy (CA) while it maps inputs that carry a trigger to a target label $y^\star$. The attack success rate (ASR) is the fraction of non-target test inputs with the trigger that are classified as $y^\star$.

\paragraph{Distributed backdoors} In canonical DBA~\cite{xie2020dba} the trigger $G$ consists of pieces $g_1,\dots,g_k$, and client $m_i$ poisons its data with $g_i$ only. Throughout this paper, \emph{distributed} refers to how malicious behavior is organized across clients, not to the communication topology. A distributed backdoor can run in server-based or decentralized FL, and we code the two separately.

\paragraph{Scope} We include studies in which the malicious training contributions of several FL clients implant a conditional targeted behavior. The clients either split trigger information among themselves or coordinate their roles, updates, or targets. Single-client attacks~\cite{bhagoji2019lens,wang2020tails,zhang2022neurotoxin,dai2023chameleon,zhang2023a3fl,nguyen2023iba,li2023threedfed,li2024darkfed} and general poisoning defenses serve as comparators. We exclude test-time perturbations without training-stage implantation, such as NI-GDBA~\cite{li2025nigdba}, and discuss them only as adjacent evidence (Appendix~\ref{app:extraction}). We focus on horizontal FL and discuss vertical FL in Section~\ref{sec:lessons}.

\paragraph{Corpus} A structured search of IEEE Xplore, Scopus, and Web of Science (cutoff 5 November 2026) returned 86 records. After deduplication, 49 remained, of which 26 were attack-relevant reports. Fifteen study lineages had complete full texts, and four more came from citation and method-led retrieval. Table~\ref{tab:coding} lists the resulting nineteen primary studies. Appendix~\ref{app:method} gives the queries, the screening procedure, and its limitations, and Appendix~\ref{app:extraction} gives the per-study extraction. Prior surveys cover FL backdoors broadly~\cite{li2025survey,wan2024survey,bellachia2026sok,djemaa2026survey,rodriguez2023threats,nguyen2024backdoorsurvey}, but none of them separates the resource effect from the composition effect or tests them. Table~\ref{tab:position} compares their organizing questions with ours.

\begin{table*}[t]
\caption{Positioning by organizing question and analytical output. The comparison describes each work's stated scope; it does not treat topic presence as a measure of analytical depth.}\label{tab:position}
\centering\scriptsize\setlength{\tabcolsep}{4pt}
\begin{tabularx}{\textwidth}{P{.18\textwidth}P{.25\textwidth}YY}
\toprule
Work & Organizing question & Principal synthesis & Focus developed here\\
\midrule
Wan et al.~\cite{wan2024survey} & Data/model poisoning in wireless FL & Attack phases and defense stages & Joint-contribution mechanism and matched exposure\\
Li et al.~\cite{li2025survey} & HFL/VFL backdoor attack and defense & Taxonomies and evaluation criteria & Component-level tests linked to study conditions\\
Bellachia et al.~\cite{bellachia2026sok} & Backdoor workflow and interventions & Lifecycle taxonomy and empirical comparisons & Mechanism-specific interpretation of coordination claims\\
Djemaa et al.~\cite{djemaa2026survey} & Poisoning under heterogeneity & Attack--defense alignment and operational assumptions & Separation of resource, exposure, and composition effects\\
This SoK & What do malicious pieces combine to achieve? & Mechanism, operating-condition and defense-test mapping & Resource versus composition effects, tested in a paired experiment; observer versus protected unit\\
\bottomrule
\end{tabularx}
\end{table*}

\section{Systematization}\label{sec:sys}
The literature uses one label, distributed backdoor, for constructions that differ along several independent axes. Table~\ref{tab:coding} codes the nineteen primary studies on these axes, and this section explains each axis.

\begin{table*}[t]
\caption{Coding of the nineteen primary studies. Pieces: what is distributed across malicious clients. Ch.: contribution channel (D, local-data poisoning; U, manipulation of returned updates; D+U, both). Participation: how controlled clients enter aggregation. Post-stop: benign-only horizon after the last malicious update (Definition~3); NR, not reported. Tier: 1a, a canonical split-trigger construction is evaluated; 1b, a coordinated multi-client backdoor without an established split-trigger relation.}\label{tab:coding}
\centering\scriptsize\setlength{\tabcolsep}{2.5pt}
\begin{tabularx}{\textwidth}{P{.095\textwidth}P{.2\textwidth}P{.03\textwidth}P{.045\textwidth}P{.19\textwidth}P{.085\textwidth}YP{.022\textwidth}}
\toprule
Study & Pieces & Ch. & Topol. & Participation & Post-stop & Defenses tested & Tier\\
\midrule
DBA~\cite{xie2020dba} & Disjoint spatial patterns of one trigger & D & Server & Forced (multi-shot); staged single-shot & Staged single-shot curves & RFA, FoolsGold & 1a\\
FCBA~\cite{liu2024fcba} & 14 subsets of 4 trigger factors & D & Server & 10 of 100 sampled; per-round attackers NR & 120 rounds & DP clipping and noise & 1a\\
SADBA~\cite{feng2025sadba} & Adaptive distributed patterns & D & Server & Population threshold; sampling text inconsistent & 50 rounds & AC/NC in objective only & 1a\\
DMBA~\cite{liu2024dmba} & Distinct trigger--target pairs & D & Server & 3 attackers; per-round NR & ASR-30 & ClippedClustering, DP-FedAvg, SignGuard & 1b\\
DBA-DFL~\cite{liu2026dbadfl} & Decomposed trigger & D & P2P & Placement in graph varied & NR & FLIP, FedGame & 1a\\
FTDBA~\cite{wang2025ftdba} & Fractal sub-triggers & D & Server & 2--10 active of 10 controlled & NR & FLTrust, FLAME, AC & 1a\\
FDBA~\cite{wang2026fdba} & Fine-grained triggers, aligned updates & D+U & Server & 10\% share; per-round NR & NR & Multi-Krum, TrMean, DP, FedRecover & 1a\\
CollaPois~\cite{collapois2025} & Shared Trojaned model & U & Server & 1\% random sampling & NR & DP, norm bound, Krum, RLR & 1b\\
TFI~\cite{wang2026structure} & Fractal perturbations, structure-aware selection & D+U & Server & 10\% random; attacker share NR & NR & Krum, DP noise & 1b\\
Graph DBA~\cite{Xu2024FedGNNBackdoors} & Local subgraph triggers & D & Server & $K=5$, $M\in\{2,3\}$ & NR & FLAME, FoolsGold, RLR & 1a\\
Opt-GDBA~\cite{yang2024optgdba} & Optimized subgraph triggers & D & Server & 50\% sampled, 20\% malicious & NR & Certified defense & 1a\\
Time Bandits~\cite{abad2024timebandits} & Input-time slices & D & Server & Participation 0.1/0.5/1 & NR & STRIP & 1a\\
Improved DBA~\cite{wang2025improved} & Density-adaptive sub-triggers, projected updates & D+U & Server & Unresolved in source & NR & Krum, RFA, FoolsGold, RLR, FLAME, CRFL & 1a\\
Coop.\ DBA~\cite{yang2022cooperative} & Split triggers, rotated/scaled updates & D+U & Server & Forced (4 of 10) & None & FLAME & 1a\\
CoBA~\cite{lyu2025coba} & Common optimized trigger, constrained models & D+U & Server & 10 of 100 random; 20\% controlled & 100 iterations & 15 defenses & 1b\\
IoT CBA~\cite{alharbi2024cba} & Sum-constrained update perturbations (split variant) & D+U & Server & Full participation; sweep to 0.3 & NR & NC, geometric median, RLR & 1b\\
SSRDBA~\cite{yang2025ssrdba} & Overlapping cyclic shares & D & Server & All selected in small settings & None & RFA, FoolsGold, clipping, CRFL & 1a\\
SATA~\cite{ai2025sata} & Split triggers with fragment association & D & Server & 10 of 50 random; 5 malicious & 100 rounds & Multi-Krum, FoolsGold & 1a\\
Dynamic\-Triggers~\cite{wang2024dynamic} & Generated 4-part triggers & D & Server & Counts NR; 1:100 ratio & 50 rounds & FoolsGold & 1a\\
\bottomrule
\end{tabularx}
\end{table*}

\subsection{What is distributed}
\begin{figure*}[t]
\centering
\includegraphics[width=\textwidth]{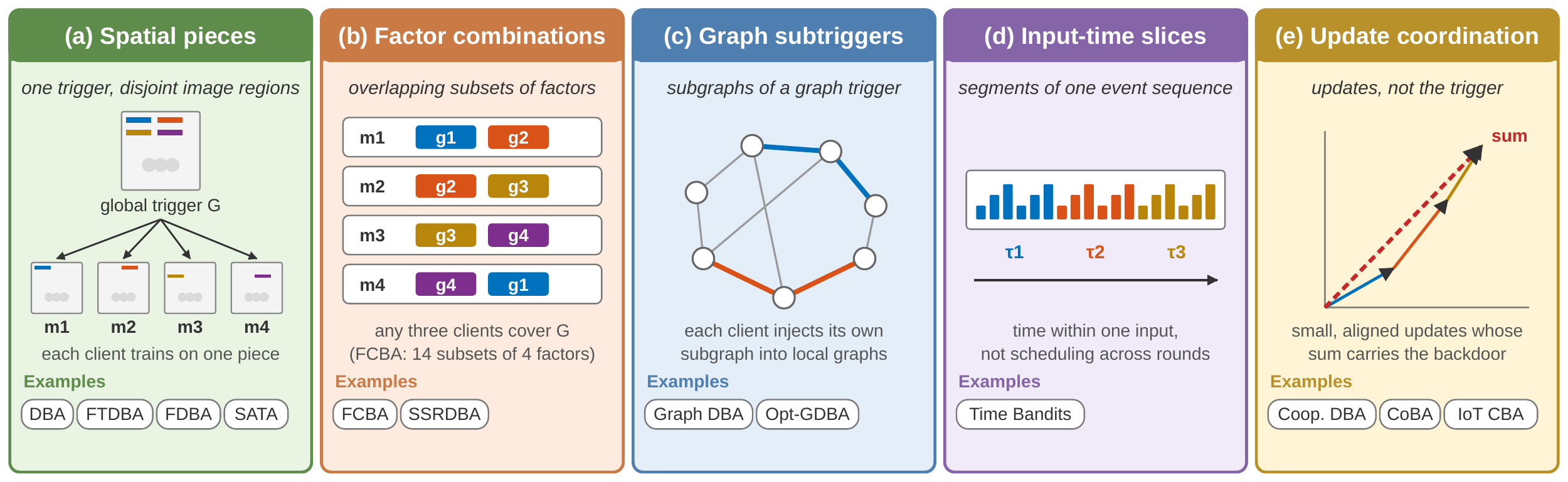}
\caption{What can a ``piece'' represent? (a)--(d) Four forms of distributed trigger information: spatial regions, overlapping combinations of factors (the cyclic shares shown are those of SSRDBA and our arm D, while FCBA assigns all 14 nonempty proper subsets of four factors), graph substructures, and time slices within one input. (e) Update-space coordination, which coordinates the submitted updates rather than the trigger and can accompany any of (a)--(d)~\cite{yang2022cooperative,lyu2025coba,alharbi2024cba}. The forms are not mutually exclusive attack classes.}\label{fig:mechanisms}
\end{figure*}

\emph{Trigger pieces.} In canonical DBA, the pieces are disjoint image patches~\cite{xie2020dba}. Later work distributes other units. FCBA assigns overlapping combinations of trigger factors, i.e., the 14 nonempty proper subsets of four factors, to 14 clients~\cite{liu2024fcba}. SSRDBA uses cyclic overlapping shares, so any $n-m+1$ clients cover the trigger~\cite{yang2025ssrdba}. Other attacks distribute subgraphs in federated graph learning~\cite{Xu2024FedGNNBackdoors,yang2024optgdba}, time bins of event streams~\cite{abad2024timebandits}, and fractal or generated sub-triggers~\cite{wang2025ftdba,wang2026fdba,wang2024dynamic,wang2025improved}. SATA lets each client associate other pieces with its own during local training~\cite{ai2025sata}.
\emph{Roles.} The shadow backdoor gives attackers different intensities rather than disjoint pieces~\cite{ren2024sba}.
\emph{Update influence.} Several attacks use one complete trigger and coordinate only the submitted updates. CerP and CoBA jointly optimize the trigger and constrain the malicious models~\cite{lyu2023cerp,lyu2025coba}. Cooperative DBA rotates and scales updates against FLAME~\cite{yang2022cooperative}. IoT CBA perturbs the updates so that their sum is preserved while each norm shrinks~\cite{alharbi2024cba}. CollaPois shares a Trojaned model~\cite{collapois2025}, and model-dependent triggers make the updates reinforce each other~\cite{gong2022coordinated}.
\emph{Targets.} DMBA and MT-FBA assign different trigger--target pairs to different clients~\cite{liu2024dmba,hao2025mtfba}.
Fig.~\ref{fig:mechanisms} illustrates these forms. Collusion therefore does not imply that the trigger is split. Table~\ref{tab:coding} marks studies without an established split-trigger relation as Tier~1b.

\subsection{How contributions reach the model}
\begin{figure}[t]
\centering
\includegraphics[width=\linewidth]{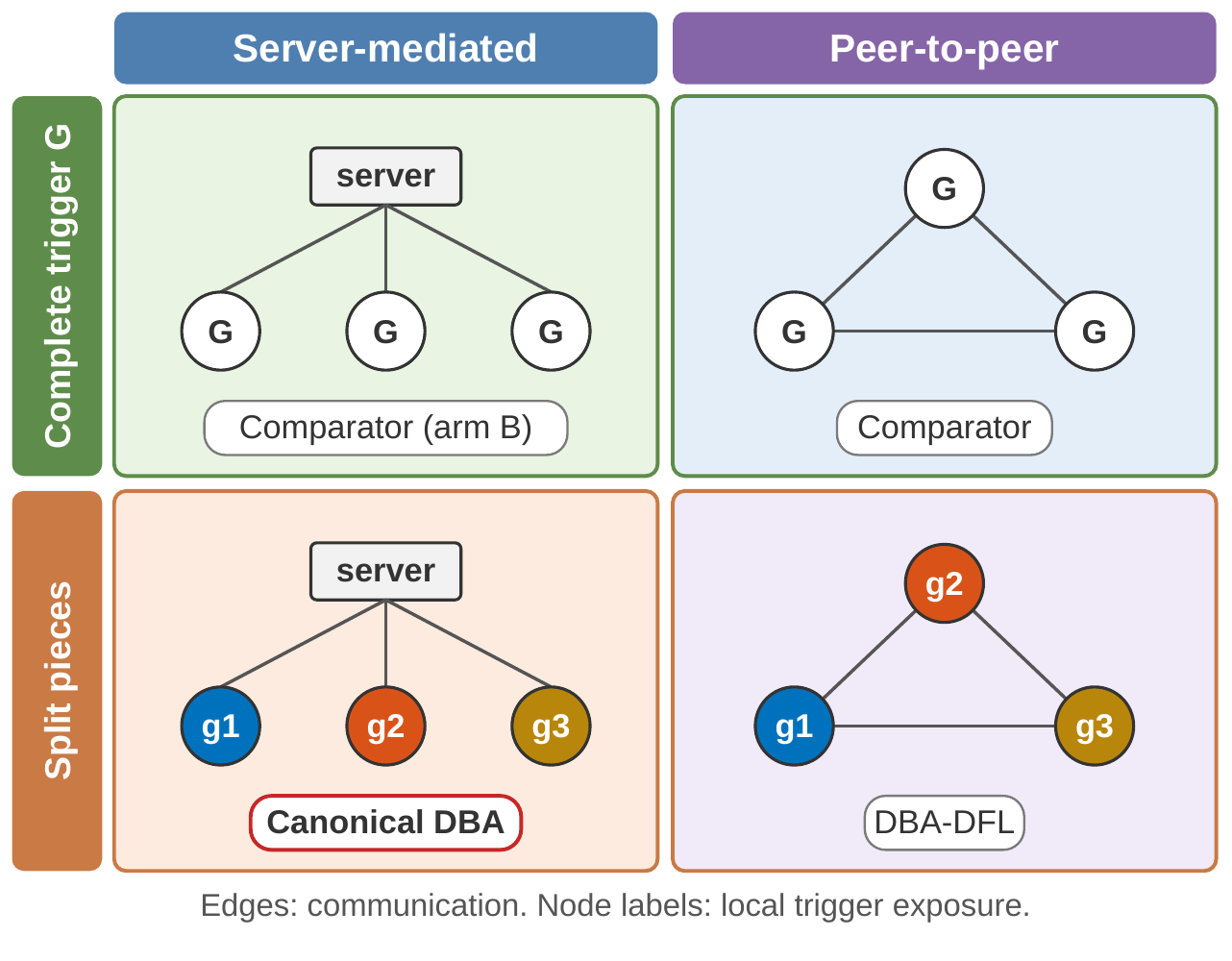}
\caption{Two independent questions: what attackers distribute (rows) and how model contributions travel (columns). Canonical DBA is the server-mediated split, and DBA-DFL studies split pieces under peer-to-peer mixing~\cite{xie2020dba,liu2026dbadfl}. The upper row is the complete-trigger comparator. Benign clients are omitted. The grid is a conceptual distinction, not a claim that every cell has equal experimental support.}\label{fig:axes}
\end{figure}

\emph{Participation.} Only sampled clients affect a round. The original DBA image experiments force all attackers into every multi-shot round~\cite{xie2020dba}, whereas most later work samples clients at random. Hence, the realized sequence of malicious updates, not the attacker fraction, determines the exposure. BackFed attributes DBA's weakness under random sampling to the loss of synchronized participation~\cite{dao2025backfed}, and early-only attacker selection can lose a time-distributed backdoor~\cite{abad2024timebandits}.
\emph{Topology.} Server-based FL and decentralized FL are coded separately (Fig.~\ref{fig:axes}). DBA-DFL~\cite{liu2026dbadfl} and peer-to-peer backdoor work~\cite{syros2023p2p} show that the placement of attackers in the communication graph changes their success. Note that graph-structured \emph{data} is a different axis from a graph \emph{topology}.
\emph{Coordination channel.} Attackers may agree offline on a piece assignment (DBA, FCBA), exchange information online each round (the malicious coordinator of IoT CBA, the joint optimization of CoBA), or not coordinate at all (Time Bandits). Online coordination needs a side channel and synchronized participation, and the attacker pays for both in a deployed network.

\subsection{Reporting practice across the corpus}
Three patterns recur in Table~\ref{tab:coding}. First, the malicious population and the delivered participation are different quantities, but most studies report only the former. DBA forces all attackers into every multi-shot round, FCBA samples 10 of 100 clients without reporting how many attackers are selected per round, and the minimum malicious percentage of SADBA is a population threshold rather than a per-round count~\cite{xie2020dba,liu2024fcba,feng2025sadba}. Second, combinatorial and redundant coverage consume resources. FCBA needs 14 malicious clients to place every nonempty proper subset of four factors, and SSRDBA matches the trigger size rather than the attacker count. Comparisons with a four-client split therefore also change the malicious population~\cite{liu2024fcba,yang2025ssrdba}. Third, persistence is a property of a schedule. FCBA retains 99.5\%, 79.5\%, and 57.8\% ASR on MNIST, CIFAR-10, and GTSRB after 120 benign rounds, CoBA reports more than 60\% after 100 normal iterations, and SATA reports 86.7\% after 100 benign rounds as the best of three runs~\cite{liu2024fcba,lyu2025coba,ai2025sata}. However, twelve of the nineteen studies report no benign-only horizon at all. Training and test pieces also need not coincide. Sundar et al.~\cite{sundar2022swift} train on several patterns and test with a smaller shared pattern, so we record the relation between training and test triggers explicitly. Several sources contain internal inconsistencies in client counts or poisoning ratios, which we retain rather than resolve (Appendix~\ref{app:extraction}).

\subsection{Definitions}\label{sec:defs}
Let $f_w$ be the global model, $\mathcal{D}$ the clean test distribution, and $x\oplus g_S$ input $x$ with the pieces $S\subseteq[k]$ applied. We write $\mathrm{ASR}_w(S)=\Pr_{(x,y)\sim\mathcal{D},\,y\neq y^\star}[f_w(x\oplus g_S)=y^\star]$.

\emph{Definition 1 (necessary composition).} For $\tau>\epsilon$, $f_w$ shows $(\epsilon,\tau)$-necessary composition if $\mathrm{ASR}_w([k])\ge\tau$ and $\max_{S\subsetneq[k]}\mathrm{ASR}_w(S)\le\epsilon$.

\emph{Definition 2 (independently effective pieces).} The pieces are independently effective if $\mathrm{ASR}_w(\{i\})\ge\tau$ for some $i$. A model with $\mathrm{ASR}_w([k])\ge\tau$ that shows neither necessary composition nor independently effective pieces shows \emph{partial} composition.

\emph{Definition 3 (post-stop persistence).} With $T_{\mathrm{last}}$ the last round that aggregates a malicious update and only benign updates afterwards, $\mathrm{ASR}\text{-}t=\mathrm{ASR}_{w_{T_{\mathrm{last}}+t}}([k])$, reported with the clean accuracy at the same round. Success measured while poisoning continues is exposure, not persistence.

\emph{Definition 4 (attacker budget).} $\mathcal{B}=(|\mathcal{M}|,U,P,\Omega,Q)$, where $U$ counts malicious updates submitted to aggregation, $P$ the poisoned samples behind them, $\Omega$ their cumulative aggregation weight, and $Q$ the total number of trigger pixels they carry. Two configurations are \emph{update-matched} when $U$, $P$, and $\Omega$ agree, and \emph{pixel-matched} when $U$ and $Q$ agree. Splitting a trigger over the same samples cannot satisfy both, because each sample then carries only part of the trigger. The \emph{resource effect} compares configurations with different $|\mathcal{M}|$ and the same trigger; the \emph{composition effect} compares a split trigger with the complete trigger under a stated matching.

\subsection{Coordination failure modes}
\begin{table*}[t]
\caption{Coordination failure modes, the attacks that address them, and the defense signals that act on them. Evidence gaps refer to the coded corpus.}\label{tab:failure-defense}
\centering\scriptsize\setlength{\tabcolsep}{4pt}
\begin{tabularx}{\textwidth}{P{.15\textwidth}P{.2\textwidth}P{.2\textwidth}YP{.2\textwidth}}
\toprule
Failure mode & Attacker's problem & Attacks addressing it & Defense signals acting on it & Main evidence gap\\
\midrule
Coverage & A trigger piece never reaches training (attacker unselected, absent, or uncooperative) & SSRDBA (overlapping shares), SATA (fragment association), FCBA (factor combinations)~\cite{yang2025ssrdba,ai2025sata,liu2024fcba} & Behavioral and inversion-based checks that search for any effective trigger, e.g.\ Neural Cleanse, FLIP, BackdoorIndicator~\cite{wang2019neuralcleanse,zhang2023flip,li2024backdoorindicator} & Redundant shares not tested against inversion-based defenses; coverage under random sampling rarely isolated\\
Detectability & Individual malicious updates are anomalous to a per-update observer & Cooperative DBA, CoBA, IoT CBA, CollaPois~\cite{yang2022cooperative,lyu2025coba,alharbi2024cba,collapois2025}; single-client analogues F3BA, 3DFed~\cite{fang2023f3ba,li2023threedfed} & Update geometry and robust aggregation (Krum, TrMean, RFA, FLAME, DeepSight, FreqFed, FoolsGold), trust anchors (FLTrust) & Aggregate-level or behavior-level checks against sum-preserving collusion are largely untested; secure aggregation hides per-update signals\\
Retention & The backdoor decays once malicious updates stop & FCBA, SATA, CoBA, PBA~\cite{liu2024fcba,ai2025sata,lyu2025coba,yang2026pba}; single-client analogues Neurotoxin, Chameleon, IBA~\cite{zhang2022neurotoxin,dai2023chameleon,nguyen2023iba} & Clipping with noise, certified smoothing (CRFL), post-training repair (fine-pruning, FedRecover)~\cite{Xie2021CRFL,liu2018fineprune,cao2023fedrecover} & No matched comparison with single-client durability methods; repair not evaluated at piece or subset level\\
Interference & Several trigger--target mappings or uncooperative clients compete for the same model capacity & DMBA, MT-FBA (distinct targets), SSRDBA (uncooperative clients)~\cite{liu2024dmba,hao2025mtfba,yang2025ssrdba} & The same update- and behavior-level signals as above; none is designed for several coordinated targets & Interference rarely measured against a single-target baseline with the same attackers\\
\bottomrule
\end{tabularx}
\end{table*}

Each construction in Table~\ref{tab:coding} addresses a specific weakness of a disjoint split. \emph{Coverage:} a piece never reaches training because its client is not sampled or does not cooperate. Overlapping shares and fragment association address this failure~\cite{yang2025ssrdba,ai2025sata,liu2024fcba}. \emph{Detectability:} individual malicious updates look anomalous to a per-update observer. Update-space coordination addresses it~\cite{yang2022cooperative,lyu2025coba,alharbi2024cba,collapois2025}, and so do single-client adaptive attacks~\cite{fang2023f3ba,li2023threedfed}. \emph{Retention:} the backdoor decays after poisoning stops. Persistence-oriented training addresses it~\cite{liu2024fcba,ai2025sata,lyu2025coba,yang2026pba}, and so do single-client durability methods~\cite{zhang2022neurotoxin,dai2023chameleon,nguyen2023iba}. \emph{Interference:} several trigger--target mappings or uncooperative clients compete for the same model capacity~\cite{liu2024dmba,hao2025mtfba,yang2025ssrdba}. Table~\ref{tab:failure-defense} maps each failure mode to the attacks that address it and to the defense signals that act on it. Evidence for one failure mode does not transfer to another. For example, a rotation attack tested with guaranteed attacker selection says nothing about coverage, and a redundant trigger does not by itself evade an update filter.

\section{What the Literature Establishes}\label{sec:evidence}
Table~\ref{tab:claims} maps the advantages claimed for distributed backdoors to the comparison that would support each claim and to what the coded studies report.

\begin{table*}[t]
\caption{Claims audit. For each claimed advantage of distributed backdoors, the comparison that isolates it and what the nineteen coded studies and related work provide.}\label{tab:claims}
\centering\scriptsize\setlength{\tabcolsep}{3pt}
\begin{tabularx}{\textwidth}{P{.15\textwidth}P{.25\textwidth}YP{.11\textwidth}}
\toprule
Claim & Comparison that isolates it & What the literature provides & Status\\
\midrule
Splitting makes the attack stronger & Split versus complete trigger with the same attackers, updates, and samples (update-matched) or the same trigger pixels (pixel-matched) & DBA matches total pixels but compares four attackers with one~\cite{xie2020dba}; FCBA matches pixels against DBA~\cite{liu2024fcba}; DBA-DFL and SSRDBA match pixels or trigger size~\cite{liu2026dbadfl,yang2025ssrdba}. No coded study varies attackers and split independently. BackFed, with equal attacker fractions, finds the split weaker~\cite{dao2025backfed} & Confounded with resources\\
The model learns a composite trigger & Piece-level and subset-level ASR (Definitions~1--2) & DBA reports per-piece ASR below the full trigger~\cite{xie2020dba}; Graph DBA reports local-trigger ASR, but against one centralized attacker~\cite{Xu2024FedGNNBackdoors}; no coded study reports subset tests & Untested with equal attackers\\
Distributed backdoors persist longer & Matched schedule, then benign-only rounds (Definition~3), against single-client durable attacks & 7 of 19 studies report a post-stop horizon (Table~\ref{tab:coding}); in the extracted conditions none compares with Neurotoxin, Chameleon, or IBA under a matched schedule & Partly shown, source unclear\\
Splitting evades defenses & Same defense, same budget, split versus complete trigger; acceptance rates & DBA compares split and single-attacker triggers under RFA and FoolsGold at matched pixels but unequal attackers~\cite{xie2020dba}; other studies test only their own construction (Table~\ref{tab:defense}) & Confounded with resources\\
A small attacker fraction suffices & Realized per-round participation and aggregation weight & Nominal fractions are reported; realized participation rarely is & Underspecified\\
Topology changes success & Varied placement with fixed mixing protocol & DBA-DFL and peer-to-peer work vary placement~\cite{liu2026dbadfl,syros2023p2p} & Shown for tested topologies\\
\bottomrule
\end{tabularx}
\end{table*}

Two observations follow from Table~\ref{tab:claims}. First, the comparison behind the central claim is confounded. When several attackers with pieces are compared with one attacker with the full trigger, the distributed attack also delivers more malicious updates, poisoned samples, and aggregation weight. The DBA paper itself notes that its appendix comparison, with an equal number of injections and a reduced poisoning ratio, is not fully fair, and this comparison gives mixed results across datasets~\cite{xie2020dba}. Studies that match pixels do not match attackers. BackFed matches attackers but not the per-sample trigger. Graph DBA comes closest to a piece-level test: it reports that local triggers can be as effective as the global trigger, even when a single centralized attacker implants the global trigger~\cite{Xu2024FedGNNBackdoors}. Its centralized baseline, however, uses one client against two or three distributed attackers. None of these designs isolates what splitting contributes.

Second, the success measures are often not the ones that the claims need. A high ASR during repeated poisoning is exposure rather than persistence, and twelve of the nineteen studies report no benign-only horizon. A low per-update detection rate under guaranteed attacker selection says nothing about coverage under random sampling. Strong single-client attacks~\cite{wang2020tails,zhang2022neurotoxin,zhang2023a3fl,nguyen2023iba} set the bar that any coordination benefit must clear, but they are rarely included as baselines.

\begin{figure}[t]
\centering
\includegraphics[width=\linewidth]{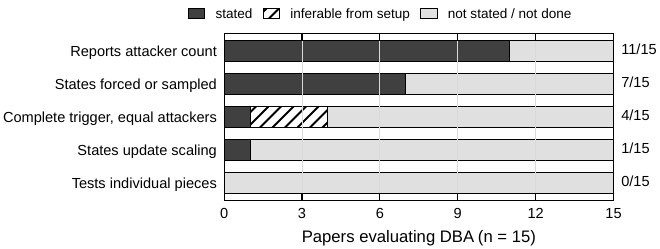}
\caption{How fifteen peer-reviewed defense and attack papers that evaluate DBA configure and measure it (Appendix~\ref{app:downstream}). Inferable: an equal-attacker complete-trigger attack is present in the setup but not presented as a comparison with DBA.}\label{fig:downstream}
\end{figure}

\paragraph{Downstream use} The belief also shapes how defenses are evaluated, because DBA is a standard attack in defense papers. We coded a purposive sample (not a census) of fifteen peer-reviewed papers from 2021 to 2025 that run DBA as a baseline: eleven defenses and four attacks, published at USENIX Security, IEEE S\&P, NDSS, NeurIPS, ICML, ICLR, CVPR, and AAAI~\cite{ozdayi2021rlr,Nguyen2022FLAME,rieger2022deepsight,fang2023f3ba,zhang2023a3fl,nguyen2023iba,huang2023lockdown,jia2023fedgame,fereidooni2024freqfed,qin2024snowball,kabir2024flshield,li2024backdoorindicator,zhuang2024lp,xie2024fedredefense,xu2025alignins} (Fig.~\ref{fig:downstream}; Appendix~\ref{app:downstream}). Four papers do not report how many clients run DBA, eight do not say whether these clients are forced into rounds or sampled, and only one states whether DBA's updates are scaled. Only one paper runs a complete-trigger attack with the same attackers by design, and three more have one implicitly in their setup. None presents it as a comparison with DBA. No paper reports whether its defense leaves individual pieces effective, because every evaluation measures only the full trigger.

Section~\ref{sec:exp} tests the first two rows of Table~\ref{tab:claims} directly, and the fourth row for the defenses that we run.

\finding{1}{No coded study compares a split trigger with a complete trigger delivered by the same attackers. Only DBA and Graph DBA report piece-level ASR, both against fewer complete-trigger attackers, and 12 of the 19 studies report no benign-only horizon. The central claims, i.e., that splitting makes a backdoor stronger, more persistent, and stealthier, are therefore confounded with attacker resources or untested. Downstream, none of 15 papers that evaluate defenses or attacks against DBA tests individual pieces.}

\section{Defenses: Who Observes What}\label{sec:defenses}
A defense observes some signal and protects some unit, and the two need not coincide. Server-side filters observe individual updates and protect the global model. Trust-anchored methods compare each update with a reference computed on clean root data~\cite{Cao2021FLTrust}. Behavior-based methods inspect models or inputs~\cite{rieger2022deepsight,wang2019neuralcleanse,gao2019strip}. Clipping and smoothing bound the influence of each update~\cite{sun2019normbound,mcmahan2018dpfedavg,Xie2021CRFL}, and recovery or pruning repairs a model after the fact~\cite{cao2023fedrecover,liu2018fineprune,wu2020fedprune}. Table~\ref{tab:defense} lists representative defenses, the signal each one needs, and the distributed constructions it has been evaluated against, either in its own paper or in later attack papers.

\begin{table*}[t]
\caption{Defense signals and evaluated constructions, including tests reported in later attack papers. A row identifies a tested pairing, not that the defense succeeded. Tier as in Table~\ref{tab:coding}; Tier 2 is adjacent evidence only.}\label{tab:defense}
\centering\scriptsize\setlength{\tabcolsep}{3pt}
\begin{tabularx}{\textwidth}{P{.2\textwidth}P{.22\textwidth}P{.17\textwidth}YP{.025\textwidth}}
\toprule
Defense & Signal and intervention & Observer and resource & Evaluated distributed construction & Tier\\
\midrule
Krum/Multi-Krum~\cite{blanchard2017krum}, TrMean~\cite{yin2018trimmed}, RFA~\cite{pillutla2022rfa} & Distance- or coordinate-based robust aggregation & Server; individual updates & DBA (RFA) in DBA; Improved DBA, FDBA, SATA, CollaPois in later attack papers & 1a\\
FLAME~\cite{Nguyen2022FLAME} & Clustering, clipping, calibrated noise & Server; individual updates & Canonical DBA in original evaluation; Cooperative DBA, FTDBA & 1a\\
FLD~\cite{Lin2023FLD} & Layer-wise model filtering & Server; individual models/updates & Canonical DBA in original evaluation & 1a\\
BoBa~\cite{jiang2026boba} & Distribution inference, overlapping clustering, voting & Server; individual updates & DBA included in original evaluation & 1a\\
FoolsGold~\cite{Fung2020FoolsGold} & Similarity of contribution histories & Server; linked histories over rounds & Sybil poisoning originally; DBA multi-shot in DBA; graph DBA; SSRDBA, SATA & 1a\\
FLIP~\cite{zhang2023flip}, FedGame~\cite{jia2023fedgame} & Trigger inversion and hardening; game-theoretic reweighting & Server/clients; local inversion & Split trigger in decentralized FL (DBA-DFL) & 1a\\
FLTrust~\cite{Cao2021FLTrust} & Root-update trust and magnitude normalization & Server; clean root dataset & Targeted scaling backdoor originally; FTDBA and CoBA in later attack papers & 1a\\
CRFL~\cite{Xie2021CRFL} & Clipping and randomized smoothing & Server; controlled updates, specified certification conditions & Coordinated model replacement; SSRDBA, Improved DBA & 1a\\
DP clipping/noise~\cite{mcmahan2018dpfedavg,sun2019normbound} & Norm bound and Gaussian noise & Server; individual updates & FCBA, DMBA, TFI, CollaPois & 1a\\
FedRecover~\cite{cao2023fedrecover} & Model recovery from stored history & Server; historical updates & FDBA & 1a\\
SignGuard~\cite{xu2022signguard}, ClippedClustering~\cite{li2023clippedclustering} & Sign statistics; clipping and clustering & Server; individual updates & DMBA (multi-target) & 1b\\
CrowdGuard~\cite{Rieger2024CrowdGuard} & Client feedback and hidden-neuron behavior & Clients; feedback and trusted execution & Adaptive and multiple backdoors; distinct from split-trigger construction & 2\\
FreqFed~\cite{fereidooni2024freqfed}, BackdoorIndicator~\cite{li2024backdoorindicator} & Frequency analysis of updates; OOD indicator tasks & Server; individual models & DBA in original evaluation & 1a\\
DeepSight~\cite{rieger2022deepsight} & Model inspection and clustering & Server; individual models & DBA in original evaluation & 1a\\
FLDetector~\cite{zhang2022fldetector}, MESAS~\cite{krauss2023mesas} & Update consistency over rounds; multi-metric statistics & Server; update histories & No split-trigger evaluation identified & 2\\
Fine-pruning~\cite{liu2018fineprune,wu2020fedprune} & Post-training pruning and fine-tuning & Server and/or clients; activations & Single-trigger backdoors; no split-trigger test identified & 2\\
\bottomrule
\end{tabularx}
\end{table*}

The observer matters because coordination targets it. Some constructions constrain the \emph{sum} of the malicious updates while they shrink or rotate each one~\cite{alharbi2024cba,yang2022cooperative}. They defeat exactly the per-update signal that most server-side filters use. Secure aggregation~\cite{bonawitz2017} removes this signal altogether, because it reveals only the aggregate. Defenses that act on the aggregate or on model behavior are less exposed to this strategy. However, they cost clean utility or need trusted resources, e.g., root data or trusted execution~\cite{Rieger2024CrowdGuard}. For split triggers, behavior-based repair raises a specific question. Removing or inverting the composite trigger may leave individually effective pieces active, so a repair should be evaluated at the piece and subset level. No coded study reports such an evaluation, and only the attacking papers themselves evaluate defenses against redundant shares. We run this evaluation for three repairs in Section~\ref{sec:exp-defense}.

\section{Disentangling Resources from Composition}\label{sec:exp}
The claims audit leaves three questions open, because no coded study varies the number of attackers and the split independently (Finding~1).
\begin{itemize}
\item \textbf{RQ1.} Where does the reported advantage of distributed backdoors come from: the additional attackers, or the split?
\item \textbf{RQ2.} What does a model trained on pieces learn: one composite trigger, or several separate ones?
\item \textbf{RQ3.} When does splitting help?
\end{itemize}
We answer them with a paired design. All arms share the same clients, data, and sampling sequence, and each contrast changes one design choice: the number of attackers or the allocation of the trigger. A new allocation also changes the trigger signal per sample, so each contrast compares two allocation strategies under a stated matching (Definition~4).

\begin{table}[t]
\caption{Arms and what they hold fixed. Budget components (Definition~4) are relative to arm B; for A they are expected values under random participation; $\approx$: matched to within 4\%. Rows below the line use 14 attackers and are relative to B14. Supplementary settings reuse these arms: DBA's scaled single-shot setting (A, B, C, and A4, one attacker injecting four times), pieces moved apart, to bars 8 pixels apart or to the image corners (B, C, B-q4), a Neurotoxin single attacker (A-NT), three further datasets (A, B, C), server-side repairs and larger $E_{\mathrm{adv}}$ (B, C).}\label{tab:emp-design}
\centering\scriptsize\setlength{\tabcolsep}{3pt}
\begin{tabular}{lclccccl}
\toprule
Arm & $|\mathcal{M}|$ & Trigger per sample & $U$ & $P$ & $\Omega$ & $Q$ & Isolates\\
\midrule
A & 1 & $G$ & 1/4 & 1/4 & 1/4 & 1/4 & resource\\
B & 4 & $G$ & 1 & 1 & 1 & 1 & reference\\
C & 4 & $g_i$ & 1 & 1 & 1 & 1/4 & composition (upd.)\\
D & 4 & $g_i\cup g_{i+1}$ & 1 & 1 & 1 & 1/2 & redundancy\\
B-q4 & 4 & $G$ & 1 & 1/4 & 1 & $\approx$1/4 & composition (pix.)\\
B-q2 & 4 & $G$ & 1 & 1/2 & 1 & $\approx$1/2 & redundancy (pix.)\\
\midrule
B14 & 14 & $G$ & 1 & 1 & 1 & 1 & reference\\
FCBA14 & 14 & subset $S_i\subsetneq G$ & 1 & 1 & 1 & 1/2 & composition (upd.)\\
B14-q2 & 14 & $G$ & 1 & 1/2 & 1 & $\approx$1/2 & composition (pix.)\\
\bottomrule
\end{tabular}
\end{table}

\begin{figure*}[t]
\centering
\includegraphics[width=0.5\textwidth]{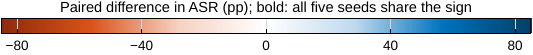}\\[-2pt]
\subfloat[End of the attack window]{\includegraphics[width=0.49\textwidth]{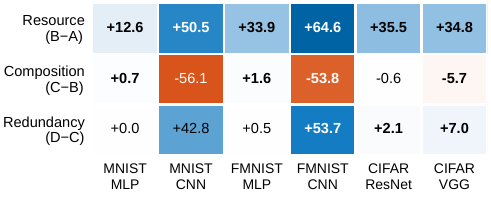}\label{fig:emp-effects-a}}\hfill
\subfloat[After 100 benign rounds]{\includegraphics[width=0.49\textwidth]{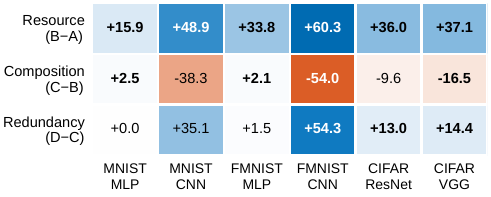}\label{fig:emp-effects-b}}
\caption{Resource, composition, and redundancy effects under random participation without defense. Each cell gives the mean paired difference over five seeds in percentage points. Bold: all five seeds share the sign. Adding attackers (B$-$A) helps in every configuration. Splitting at an equal update and sample budget (C$-$B) helps only slightly on the near-saturated MLPs, and it hurts on the CNNs at the calibrated local training (but see Fig.~\ref{fig:emp-sens-a}). Numerical values are in Table~\ref{tab:emp-paired}.}\label{fig:emp-effects}
\end{figure*}

\begin{figure*}[t]
\centering
\includegraphics[width=0.98\textwidth]{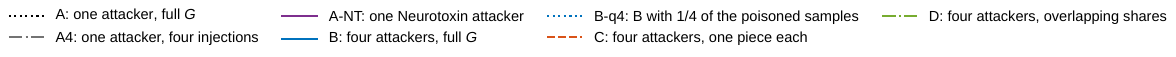}\\[-2pt]
\subfloat[Compact]{\includegraphics[width=0.245\textwidth]{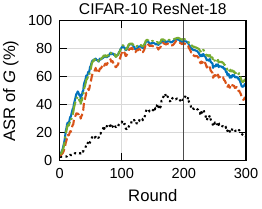}}\hfill
\subfloat[Compact]{\includegraphics[width=0.245\textwidth]{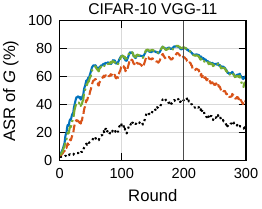}}\hfill
\subfloat[Compact]{\includegraphics[width=0.245\textwidth]{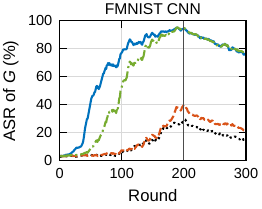}}\hfill
\subfloat[Compact]{\includegraphics[width=0.245\textwidth]{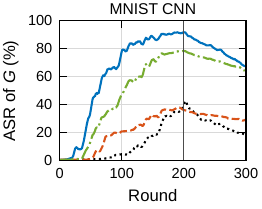}}\\[-4pt]
\subfloat[DBA-style setting]{\includegraphics[width=0.245\textwidth]{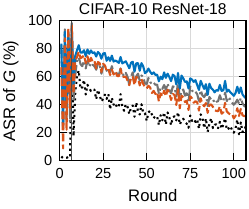}\label{fig:emp-traj-dba}}\hfill
\subfloat[Corners]{\includegraphics[width=0.245\textwidth]{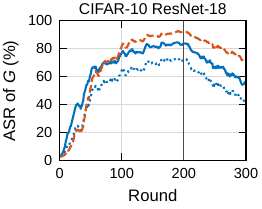}\label{fig:emp-traj-corners}}\hfill
\subfloat[Corners]{\includegraphics[width=0.245\textwidth]{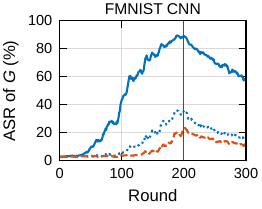}}\hfill
\subfloat[Neurotoxin]{\includegraphics[width=0.245\textwidth]{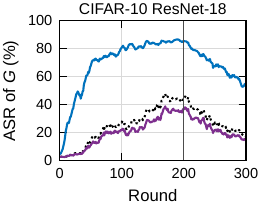}\label{fig:emp-traj-nt}}
\caption{ASR of the full trigger per round (mean over five seeds; 2{,}000-image test subset; five-round moving average except in (e)). The vertical line marks the last round with malicious participants. Afterwards, only benign clients train. (a)--(d): main experiment with DBA's compact layout, random participation. (e): DBA-style scaled single-shot setting (partial replication), injections in rounds 1, 3, 5, and 7. (f)--(g): pieces in the four image corners. (h): a Neurotoxin single attacker against the plain single attacker and arm B.}\label{fig:emp-traj}
\end{figure*}

\subsection{Design}
\emph{Arms.} Four controlled clients $m_1,\dots,m_4$ are fixed per seed (Table~\ref{tab:emp-design}). In arm A, only $m_1$ attacks, with the full trigger $G$. In B, all four clients use $G$. In C, client $m_i$ uses only piece $g_i$, as in DBA. In D, client $m_i$ uses the overlapping share $g_i\cup g_{(i\bmod 4)+1}$, so any three clients cover $G$, as in SSRDBA. B, C, and D submit the same malicious updates from the same poisoned samples with the same aggregation weight (update-matched), and differ only in the trigger pixels per sample. Hence, B$-$A is the resource effect, C$-$B the composition effect, and D$-$C the redundancy effect. B-q4 and B-q2 apply $G$ to one quarter and one half of B's poisoned samples, which matches the trigger pixels of C and D (pixel-matched). For an FCBA-style allocation~\cite{liu2024fcba}, we use 14 controlled clients. In B14, all of them use $G$. In FCBA14, each uses a different nonempty proper subset of the four pieces, and B14-q2 is the pixel-matched counterpart.

\emph{Setting.} One hundred clients hold Dirichlet ($\alpha=0.5$) label partitions. Each round samples ten clients and aggregates them with uniform FedAvg (SGD, learning rate 0.01, momentum 0.9, batch 64, two local epochs). The main experiment covers MNIST and Fashion-MNIST (FMNIST) with a two-layer CNN and an MLP, and CIFAR-10 with ResNet-18 and VGG-11, each starting from a clean federated checkpoint. Supplementary runs add ResNet-18 on grayscale CIFAR-10, FMNIST, and GTSRB. Following DBA~\cite{xie2020dba}, $G$ consists of four one-pixel-high bars in a $2\times2$ grid in the top-left corner, with gaps of 2 pixels (3 on CIFAR-10). The bars are 4 pixels wide on MNIST and FMNIST and 6 on CIFAR-10, and the target label is 2. Each malicious batch of 64 contains 20 (MNIST, FMNIST) or 5 (CIFAR-10) triggered samples, and malicious updates are not scaled.

\emph{Participation and measurement.} Under \emph{random} participation, the main setting, the controlled clients are sampled like any other client during a 200-round window. Under \emph{forced} participation, they join every round of a 20-round window with the same expected number of malicious updates. Then they leave, and 100 benign rounds follow. We measure the clean accuracy and the ASR of all 15 nonempty piece subsets at $t\in\{0,25,50,100\}$ rounds after the window, and the ASR of $G$ in every round. Because several arms are bimodal across seeds, we report each paired difference as the mean with the per-seed range and the number of seeds that agree in sign.

\emph{Defenses.} Under random participation, we repeat B, C, and D with Multi-Krum ($f=2$), norm clipping with Gaussian noise, FLAME, and FLTrust. We also repair retrained CIFAR-10 models with three server-side methods (Section~\ref{sec:exp-defense}).

\emph{Protocol.} Seed 0 was used only for calibration, and seeds 1--5 are reported. The malicious local epochs $E_{\mathrm{adv}}$ are the smallest value in $\{2,4,6\}$ at which arm B reached 90\% ASR: 2 for MNIST and FMNIST, and 6 for CIFAR-10, where no candidate reached 90\%. All settings were fixed in a hashed configuration before the reported seeds were run. The main matrix has 510 runs, of which 490 completed. We then added 190 runs under written protocols and, after an internal review, 175 more (Table~\ref{tab:emp-design}). We report these supplementary runs as exploratory evidence (Appendix~\ref{app:experiment}).

\subsection{RQ1: Where Does the Advantage Come From?}\label{sec:rq1}
\emph{The attackers, not the split.} Adding attackers raised ASR in every seed of all 24 cells (dataset, architecture, participation, and time) of the main experiment, by 7.5 to 64.6 points on average (Fig.~\ref{fig:emp-effects}). The same holds on GTSRB (8.4 and 12.5 points at $t=0$ and $t=100$) and with 14 attackers (B14$-$A: 48.7 and 58.2 points on CIFAR-10/ResNet-18, 74.5 and 83.4 on FMNIST/CNN; Fig.~\ref{fig:emp-sens-b}). In contrast, splitting the trigger among the same attackers moves ASR much less. With DBA's compact layout on CIFAR-10, C reaches nearly the same ASR as B by the end of the window and then decays faster ($-9.6$ and $-16.5$ points at $t=100$ on ResNet-18 and VGG-11). On grayscale CIFAR-10 and GTSRB, the gap is $-7.0$ and $-1.3$ points. On the MLPs, splitting gains 0.2 to 12.5 points.

\emph{A large penalty on small CNNs is an operating-point effect.} On the MNIST and FMNIST CNNs, C trails B by up to 56.1 points, because at $E_{\mathrm{adv}}=2$ the split often fails to implant (Fig.~\ref{fig:emp-traj}). This penalty is a result of calibrating on B. With 4 or 6 local epochs, C implants in four or five of five seeds. With an equal tuning budget, i.e., the best of $E_{\mathrm{adv}}\in\{2,4,6\}$ for each arm, both arms select 6, and C matches or exceeds B ($+0.8$ and $+4.9$ points on FMNIST, $+1.7$ and $+8.7$ on MNIST at $t=0$ and $t=100$; Fig.~\ref{fig:emp-sens-a}). Part of the remaining gap is a pixel dose, because a C sample carries a quarter of B's trigger pixels. Against the pixel-matched B-q4, the compact split shows no consistent sign (Fig.~\ref{fig:emp-conditions-a}).

\emph{Combinatorial allocation adds little.} With 14 attackers holding different piece subsets, FCBA14 exceeds B14 in every seed, by 1.0 and 9.1 points on CIFAR-10/ResNet-18 and by 0.2 and 1.9 on FMNIST/CNN, with similar margins over B14-q2. This gain is an order of magnitude smaller than the resource effect of the same 14 attackers (Fig.~\ref{fig:emp-sens-b}).

\emph{DBA's own result, explained.} We reproduce DBA's single-shot setting on CIFAR-10/ResNet-18, where attackers inject scaled updates one after another in rounds 1, 3, 5, and 7. We match DBA's effective scale $\gamma\eta/n=1$ and keep our optimizer (Section~\ref{sec:exp-limits}). All arms reach 90 to 98\% ASR after the last injection (Fig.~\ref{fig:emp-traj-dba}). After 100 rounds, C retains 8.5 points more than one attacker that injects once, in all four comparable seeds, which reproduces DBA's reported advantage. However, B, with the same four attackers and the full trigger, retains 22.8 points more than that attacker, and C trails B in every seed. Moreover, a single attacker that injects four times (A4) comes within 6.1 points of B. DBA's advantage is therefore the advantage of four injections, not of four pieces. As an auxiliary check, our adaptation of Neurotoxin~\cite{zhang2022neurotoxin} did not reach the plain single attacker (Table~\ref{tab:emp-nt}), but it may underperform the tuned original.

\finding{2}{The reported advantage of distributed backdoors is mainly a budget effect. More attackers submit more poisoned updates, and this raised ASR in every configuration and seed. Splitting the same budget into pieces changed ASR much less once both arms were tuned alike, and in DBA's own setting one attacker that injects as often as four attackers comes within 6.1 points of them.}

\subsection{RQ2: What Does a Split-Trained Model Learn?}\label{sec:rq2}
\begin{table}[t]
\caption{Composition class of each model at the end of the attack window ($\tau=0.8$, $\epsilon=0.2$; ten models per configuration and arm, both participation modes). Classes are assigned in order: not implanted if $\mathrm{ASR}(G)<\tau$; independent if a single piece reaches $\tau$; necessary if every proper subset stays at or below $\epsilon$; partial otherwise. Last two columns: median ratio of the best single-piece ASR to the full-trigger ASR (models with $\mathrm{ASR}(G)\ge0.5$) for the row's arm and for arm B. Configuration rows refer to arm C.}\label{tab:emp-classes}
\centering\scriptsize\setlength{\tabcolsep}{2.2pt}
\begin{tabular}{llcccccc}
\toprule
& & \multicolumn{4}{c}{Number of models} & \multicolumn{2}{c}{Single piece / $G$}\\
\cmidrule(lr){3-6}\cmidrule(lr){7-8}
Dataset & Model & Nec. & Part. & Indep. & Not impl. & Row arm & Arm B\\
\midrule
MNIST & MLP & 0 & 2 & 8 & 0 & 0.86 & 0.04\\
MNIST & CNN & 0 & 0 & 3 & 7 & 0.94 & 0.25\\
FMNIST & MLP & 0 & 9 & 1 & 0 & 0.69 & 0.11\\
FMNIST & CNN & 0 & 3 & 2 & 5 & 0.73 & 0.16\\
CIFAR-10 & ResNet-18 & 0 & 4 & 5 & 1 & 0.88 & 0.30\\
CIFAR-10 & VGG-11 & 0 & 6 & 0 & 4 & 0.89 & 0.30\\
\midrule
\multicolumn{2}{l}{All, arm C} & 0 & 24 & 19 & 17 & 0.86 & \\
\multicolumn{2}{l}{All, arm B} & 0 & 57 & 0 & 3 & 0.19 & \\
\multicolumn{2}{l}{All, arm D} & 0 & 51 & 5 & 4 & 0.62 & \\
\bottomrule
\end{tabular}
\end{table}

\begin{figure}[t]
\centering
\includegraphics[width=\linewidth]{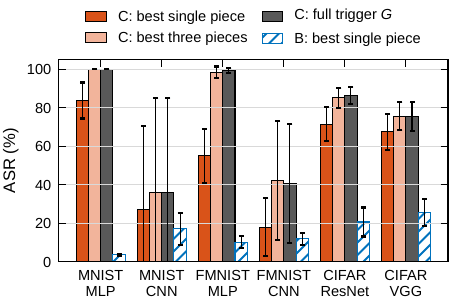}
\caption{What split training implants: ASR at the end of the attack window under random participation (mean $\pm$ one standard deviation over five seeds). Models trained on pieces (C) respond to their best single piece almost as strongly as to $G$; models trained on $G$ (B) do not.}\label{fig:emp-subsets}
\end{figure}

\emph{Pieces become triggers of their own.} The clearest difference between arms is how the model responds to single pieces. In arm C, the best single piece achieves a median 0.86 of the full-trigger ASR, and in 19 of 60 models one piece alone reaches 80\% ASR. In arm B, no single piece reaches 80\%, and the median ratio is 0.19 (Fig.~\ref{fig:emp-subsets}). Seen from the other side, the full trigger adds a median 13.2 points over the best single piece in C models, against 76.0 points in B models. The supplementary experiments agree: split-trained models respond to their best piece at a median 0.64 to 0.99 of their full-trigger ASR, and models trained on $G$ at 0.14 to 0.46, except where B is above 94\% ASR. Under our strict test for necessary composition (Definition~1), no model in any arm requires all pieces (Table~\ref{tab:emp-classes}).

\emph{Overlap buys persistence at half the pixels.} The redundancy effect D$-$C is positive in 19 of 24 cells. On CIFAR-10, D retains 10 to 14 points more than C after 100 benign rounds, about as much as B, while it carries half of B's trigger pixels. Against the pixel-matched B-q2, D is higher in every seed. Redundant shares, as in SSRDBA, thus address the retention weakness of a disjoint split.

\finding{3}{Split training produces several partially working triggers, not one composite trigger. Models trained on pieces respond to their best single piece at a median 86\% of their full-trigger ASR, against 19\% for models trained on $G$. A defense evaluated only on the full trigger therefore does not measure the backdoor that a split attack implants.}

\subsection{RQ3: When Does Splitting Help?}\label{sec:rq3}
\begin{figure*}[t]
\centering
\includegraphics[width=0.5\textwidth]{figures/heat_cbar.pdf}\\[-2pt]
\subfloat[CIFAR-10 / ResNet-18 across conditions]{\includegraphics[width=0.48\textwidth]{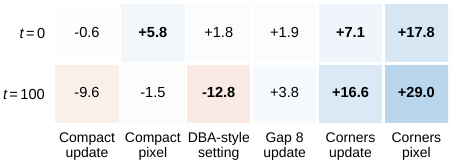}\label{fig:emp-conditions-a}}\hfill
\subfloat[Compact versus corner layout across models]{\includegraphics[width=0.48\textwidth]{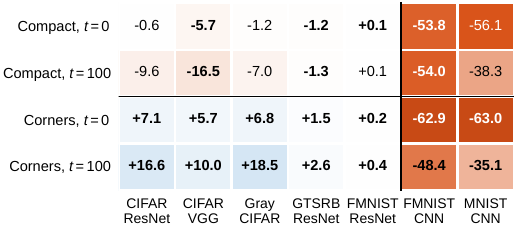}\label{fig:emp-conditions-b}}
\caption{The composition effect C$-$B depends on the layout and the data. (a) CIFAR-10/ResNet-18: DBA's compact $2\times2$ layout against B (update-matched) or B-q4 (pixel-matched), a DBA-style scaled single-shot setting, bars 8 pixels apart, and the same pieces at the four image corners. (b) Update-matched C$-$B with the compact (top) and corner (bottom) layouts. Gray CIFAR, GTSRB, and FMNIST use ResNet-18; the two CNNs (right of the line) use the calibrated $E_{\mathrm{adv}}=2$. Cells as in Fig.~\ref{fig:emp-effects}. In these exploratory settings, separating the pieces turns the effect positive on the 32$\times$32 datasets, most clearly after poisoning stops. FMNIST/ResNet-18 is saturated. Values are in Tables~\ref{tab:emp-controls} and~\ref{tab:emp-controls2}.}\label{fig:emp-conditions}
\end{figure*}

\begin{figure*}[t]
\centering
\includegraphics[width=0.5\textwidth]{figures/heat_cbar.pdf}\\[-2pt]
\subfloat[Local training, CNNs]{\includegraphics[width=0.32\textwidth]{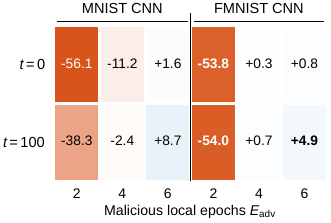}\label{fig:emp-sens-a}}\hfill
\subfloat[FCBA-style, 14 attackers]{\includegraphics[width=0.32\textwidth]{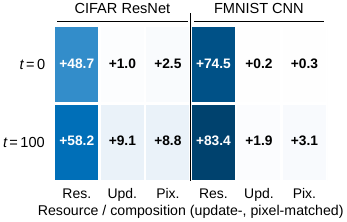}\label{fig:emp-sens-b}}\hfill
\subfloat[After Neural Cleanse, CIFAR-10]{\includegraphics[width=0.32\textwidth]{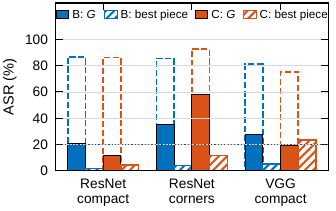}\label{fig:emp-sens-c}}
\caption{Supplementary tests. (a) C$-$B on the MNIST and FMNIST CNNs against the malicious local epochs $E_{\mathrm{adv}}$, where 2 is the calibrated value of the main experiment. (b) Resource effect B14$-$A (Res.) and composition effect of FCBA14 against B14 (Upd., update-matched) and B14-q2 (Pix., pixel-matched). In (a) and (b), cells are as in Fig.~\ref{fig:emp-effects}. (c) Mean ASR over five seeds after Neural Cleanse unlearning: $G$ (filled) and the best single piece (hatched). Dashed outlines show the ASR of $G$ before the repair, and the dotted line marks 20\%. More local training removes most of the CNN penalty, an FCBA-style split gains little next to the resource effect, and pieces fall together with $G$ under repair. Values are in Tables~\ref{tab:emp-controls2} and~\ref{tab:emp-repair}.}\label{fig:emp-sens}
\end{figure*}

\begin{table*}[t]
\caption{Defended runs under random participation (percent, mean over five seeds; standard deviations in the artifact). CA and ASR are measured at the end of the attack window; acceptance is the share of submitted malicious updates admitted to aggregation. Clip+noise always admits every update but rescales it and adds noise. CIFAR-10 rows use the corrected handling of batch-normalization statistics and $\sigma=0.0005$ (Appendix~\ref{app:experiment}).}\label{tab:emp-defense}
\centering\scriptsize\setlength{\tabcolsep}{4pt}
\resizebox{\textwidth}{!}{%
\begin{tabular}{llccccccc}
\toprule
& & & \multicolumn{3}{c}{ASR} & \multicolumn{3}{c}{Malicious updates accepted}\\
\cmidrule(lr){4-6}\cmidrule(lr){7-9}
Dataset / model & Defense & CA (B) & B & C & D & B & C & D\\
\midrule
MNIST / CNN & Multi-Krum & 99.02 & 0.16 & 0.15 & 0.15 & 0.98 & 0.98 & 0.98\\
MNIST / CNN & Clip+noise & 99.05 & 0.41 & 0.13 & 0.17 & 100.00 & 100.00 & 100.00\\
MNIST / CNN & FLAME & 99.02 & 0.12 & 0.13 & 0.12 & 19.55 & 23.13 & 22.08\\
MNIST / CNN & FLTrust & 98.51 & 0.16 & 0.15 & 0.17 & 11.93 & 10.31 & 10.01\\
FMNIST / CNN & Multi-Krum & 88.40 & 1.39 & 1.38 & 1.39 & 0.99 & 1.48 & 1.48\\
FMNIST / CNN & Clip+noise & 89.02 & 66.18 & 2.24 & 11.82 & 100.00 & 100.00 & 100.00\\
FMNIST / CNN & FLAME & 87.27 & 1.65 & 0.82 & 1.21 & 19.05 & 21.90 & 21.08\\
FMNIST / CNN & FLTrust & 87.59 & 2.70 & 2.48 & 2.52 & 20.43 & 21.27 & 20.19\\
CIFAR-10 / ResNet-18 & Multi-Krum & 84.45 & 9.39 & 3.04 & 3.45 & 5.19 & 3.62 & 3.38\\
CIFAR-10 / ResNet-18 & Clip+noise & 86.61 & 71.46 & 60.66 & 70.62 & 100.00 & 100.00 & 100.00\\
CIFAR-10 / ResNet-18 & FLAME & 81.52 & 27.66 & 8.78 & 17.76 & 21.82 & 16.57 & 17.27\\
CIFAR-10 / ResNet-18 & FLTrust & 86.46 & 20.56 & 6.62 & 9.62 & 63.54 & 58.16 & 63.98\\
\bottomrule
\end{tabular}%
}
\end{table*}
\emph{The layout decides.} DBA places the four bars within a few pixels of each other. Moving the same pieces apart turns the composition effect positive (Fig.~\ref{fig:emp-conditions}). On CIFAR-10/ResNet-18, C$-$B after 100 benign rounds rises from $-9.6$ points with the compact layout to $+3.8$ with bars 8 pixels apart and to $+16.6$ with the pieces at the four corners. In the corner layout, C beats B in every seed, and it beats the pixel-matched B-q4 by 29.0 points. The gain comes from the split: moving the pieces barely changes the complete trigger ($+2.3$ points) but raises the split one by 28.5 points, and the gain grows after poisoning stops (Fig.~\ref{fig:emp-traj-corners}).

\emph{The corner gain generalizes across models and data.} VGG-11 shows the same effect, with C ahead of B in every seed (5.8 and 10.0 points at $t=0$ and $t=100$, against $-5.7$ and $-16.5$ in the compact layout). The gain does not depend on color or on the classes of CIFAR-10. On grayscale CIFAR-10, C beats B in every seed by 6.8 and 18.5 points. On GTSRB, where the compact split trails B, the corner split beats it in every seed (1.5 and 2.6 points). FMNIST on ResNet-18 is saturated above 99\% and cannot separate the layouts. On the MNIST and FMNIST CNNs, the corner split also suffers the operating-point penalty of RQ1 at $E_{\mathrm{adv}}=2$ (C$-$B of about $-63$ points), and we did not retest it with more local epochs.

\finding{4}{Splitting helps when the pieces are placed apart. With the pieces at the image corners, the split trigger beat the complete trigger in every seed on CIFAR-10 with two architectures, on grayscale CIFAR-10, and on GTSRB, and its advantage grew after poisoning stopped. DBA's compact layout, which the literature has used as the default, is the layout in which splitting does not help.}

\subsection{Defenses}\label{sec:exp-defense}
\emph{Splitting does not evade aggregation defenses.} On MNIST and FMNIST, Multi-Krum, FLAME, and FLTrust reduce every arm below 3\% ASR (Table~\ref{tab:emp-defense}). On CIFAR-10/ResNet-18, they leave B at 9.4 to 27.7\% ASR but C at only 3.0 to 8.8\%, and they admit C's updates at the same or a lower rate than B's (e.g., FLTrust 58.2\% versus 63.5\%). Clipping with noise that costs less than one point of clean accuracy stops neither arm on CIFAR-10 (B 71.5\%, C 60.7\%).

\emph{Corner-split backdoors survive repair better.} Finding~3 raises a concrete risk: a repair that removes the full trigger might leave a working piece. To test it, we retrained B and C on CIFAR-10 (ResNet-18 in the compact and corner layouts, VGG-11 in the compact layout) and repaired each model with fine-pruning~\cite{liu2018fineprune}, clean fine-tuning, and Neural Cleanse~\cite{wang2019neuralcleanse}, all with the 100 root images of FLTrust (Table~\ref{tab:emp-repair}). Fine-pruning and fine-tuning left the backdoors in place. Neural Cleanse lowered ASR($G$) to 12 to 58\% at a cost of 5.6 to 9.6 points of clean accuracy, and it suppressed the pieces too (Fig.~\ref{fig:emp-sens-c}). No model kept a working piece after losing the full trigger. The layout, however, again made the difference. In the corner layout, the split backdoor kept 58.4\% ASR after Neural Cleanse against 35.4\% for the complete trigger, ahead in every seed by 13.0 to 47.7 points, although it started only 7.1 points ahead.

\finding{5}{Splitting did not make malicious updates harder to filter, and Neural Cleanse suppressed single pieces together with the full trigger. In the corner layout, however, the split backdoor survived the repair far better than the complete one. Defense evaluations should therefore test layouts other than DBA's default.}

\subsection{Limitations}\label{sec:exp-limits}
\emph{Scope and statistics.} The experiments test spatial patch pieces in image classification with FedAvg, with four pieces and four or 14 attackers. Graph and temporal triggers, update-space coordination, and adaptive optimization are not tested. Each configuration has five paired seeds. For one prespecified contrast, five seeds of the same sign correspond to a two-sided sign-test $p$-value of 0.0625, and cells are not independent, so we report ranges and sign counts rather than significance.

\emph{Calibration and matching.} $E_{\mathrm{adv}}$ was calibrated on B, which favored B on the small CNNs (RQ1), and arm A was not retuned. Neither update nor pixel matching equalizes the position and shape of the trigger, and the corner layout changes the separation and the absolute position of the pieces together.

\emph{Post hoc supplements and adapted attacks.} All supplementary settings were added after the main results were known. The repair test uses 100 root images and one configuration per method. The DBA-style setting matches DBA's effective scale but not all of its hyperparameters, and our Neurotoxin attacker estimates the benign direction from the previous global update rather than from benign data.

\emph{Implementation correction.} Our first implementation passed batch-normalization statistics through the defenses, which broke the CIFAR-10 defense runs. We corrected it and repeated all CIFAR-10 defense runs, and Table~\ref{tab:emp-defense} reports the corrected runs. Appendix~\ref{app:experiment} lists all deviations from the protocol.

\section{Lessons and Open Problems}\label{sec:lessons}
Our results suggest a different picture of distributed backdoors. Their reported strength comes mainly from the attackers' budget, splitting adds a layout-dependent effect on top of it, and the implanted backdoor consists of several partially working triggers. Current evaluations cannot see this picture, because they vary the attacker resources, the trigger allocation, the realized participation, and the unit that a defense observes together. Table~\ref{tab:checklist} condenses our recommendations into a minimum reporting checklist. R1--R5 address attack evaluations, and R6 addresses defense evaluations.

\paragraph{What this means for attacks} A new distributed attack should be judged against the same attackers with the full trigger, because this baseline already captures most of the gain that the literature attributes to splitting. The design choices that do change the outcome are the ones the literature rarely reports: where the pieces are placed, whether they overlap, and how long the attackers train locally. In our experiments, corner placement gave the split a lasting advantage, and overlapping shares gave it the persistence of the complete trigger at half the pixels.

\paragraph{What this means for defenses} A defense that is evaluated against DBA with the compact layout and the full trigger is evaluated against the weakest form of the split. Split-trained models respond to single pieces, and the corner-split backdoor resisted trigger reverse engineering better than the complete one. Defense papers should therefore report piece-level results and test at least one layout with separated pieces.

\begin{table}[t]
\caption{Minimum reporting checklist for multi-client backdoor evaluations.}\label{tab:checklist}
\centering\scriptsize\setlength{\tabcolsep}{3pt}
\begin{tabular}{p{0.36\linewidth}p{0.56\linewidth}}
\toprule
Report & Why \\
\midrule
Attacker count and realized per-round participation & Separates resources from organization (R1, R4) \\
Poisoning budget $(U,P,\Omega,Q)$ and the stated matching & Makes split and complete triggers comparable (Definition~4) \\
Complete-trigger baseline with the same attackers and layout & Isolates the composition effect (R1, R3) \\
Piece layout and malicious local training & Fixes the operating point of the comparison (R3) \\
ASR of single pieces and subsets & Tests what the model learned (R2) \\
ASR after a benign-only horizon & Distinguishes persistence from exposure (R4) \\
Clean accuracy at every reported time & Rules out utility collapse \\
Malicious updates admitted by the defense & Shows what the defense actually observed \\
Piece-level ASR after a defense or repair & Checks that no piece remains effective (R6) \\
\bottomrule
\end{tabular}
\end{table}

\rec{R1 (Matched baseline).}{A claim that splitting or coordination helps needs a complete-trigger baseline with the same attackers and the same piece layout. The matching should be stated as update-matched or pixel-matched (Definition~4). Without this baseline, a resource effect of tens of percentage points is attributed to the split (Finding~2).}
\rec{R2 (Pieces and subsets).}{Report the ASR of single pieces and of subsets (Definitions~1--2). A composite trigger is a claim about what the model learned, so it needs subset evidence. In our experiments, no model met Definition~1, including the models trained on the full trigger (Finding~3).}
\rec{R3 (Layout and operating point).}{State where the pieces are placed, how far apart they are, and how long the attackers train locally. The sign of the composition effect changed with the layout (Finding~4), and its size changed with the malicious local epochs (Section~\ref{sec:rq1}). A result for one layout or operating point therefore does not transfer to another, and calibrating on one arm can favor that arm.}
\rec{R4 (Exposure and persistence).}{Report the realized per-round participation rather than a nominal attacker fraction, and a benign-only horizon after the last malicious update (Definition~3). Success measured while poisoning continues is exposure, not persistence.}
\rec{R5 (Strong baselines and seeds).}{Include strong single-client attacks~\cite{zhang2022neurotoxin,zhang2023a3fl,nguyen2023iba} and report paired per-seed outcomes. Split-trigger attacks often either implant or fail, so a mean alone can mislead.}
\rec{R6 (Defenses).}{Defense evaluations against split-trigger attacks should report the ASR of single pieces and subsets, not only of the composite pattern. None of the 15 downstream papers did this. Because split-trained models respond to single pieces, a defense that suppresses the composite trigger could leave a piece effective. This did not happen in our repair test, but the survival of the split backdoor depended on the layout (Finding~5), and only a piece-level and layout-aware evaluation shows this. Per-update filters should also be evaluated against coordinated constructions that preserve the aggregate while they shrink each update~\cite{alharbi2024cba,yang2022cooperative}.}

\paragraph{Open problems} The composition effect is still untested under adaptive optimization and for non-patch and semantic triggers. Our separated-piece results call for a systematic map of when splitting helps, over piece separation, local training, model, and data. It is also open whether the separation or the absolute position of the pieces drives the corner gain, and why corner-placed pieces resist trigger reverse engineering better than a compact complete trigger. Beyond horizontal FL, the same question arises for feature-space triggers split across parties in vertical FL~\cite{naseri2024badvfl}, for pieces that meet only in hierarchical client--edge--cloud aggregation~\cite{Liu2020ClientEdgeCloudHierFL}, and for textual trigger components in federated instruction tuning~\cite{zhao2026signal,kim2025robust}.

\section{Ethics and Potential Harms}\label{sec:ethics}
This work studies published attack constructions and does not introduce a new attack. All experiments run in simulation on public image datasets (MNIST, Fashion-MNIST, CIFAR-10, GTSRB). They involve no deployed system, no human subjects, and no personal data. The main risk of publication is that our findings guide attackers toward more effective configurations, e.g., by showing that adding malicious clients matters more than splitting a trigger. We consider this risk small, because it is well known that more malicious clients strengthen poisoning, and we provide nothing beyond reimplementations of published attacks. The benefits go mainly to defenders and evaluators: a more accurate threat model, the observation that split-trigger backdoors respond to individual pieces, and reporting requirements against overstated claims. Our separated-piece result identifies a condition under which splitting helps an attacker. We report it because defenders need it to evaluate split-trigger attacks, and it changes only the placement of the published DBA pieces. The released artifact contains the reimplemented attacks only at the level of detail already given in the original papers.

\section{Conclusion}\label{sec:conclusion}
Distributed backdoor attacks have been credited with an advantage that comes mainly from their extra attackers. By adding a complete-trigger baseline with the same attackers, we separated the two effects and found that the budget, not the split, drives the reported gains. Splitting is still not irrelevant. It turns the backdoor into several partially working triggers, and with the pieces placed at the image corners it beats the complete trigger and resists Neural Cleanse better. A distributed backdoor is therefore best understood as a resource attack whose split adds a layout-dependent effect. Attack and defense evaluations should report the matching, the layout, the pieces, and the participation on which their claims depend, and defenses should be tested against layouts other than DBA's default.

\section*{Acknowledgment}
This work was supported by JSPS KAKENHI Grant Number JP26K21220.

\emph{Use of AI systems.} We used OpenAI Codex to assist in reimplementing published attacks and defenses and in running the experiments of Section~\ref{sec:exp}, following protocols written by the authors. The authors reviewed the code, its unit tests, and the run logs. We used Anthropic Claude to assist in drawing the figures and in editing the text. As described in Appendix~\ref{app:method}, LLM-based assistants also supported literature screening and data extraction, and the authors verified every inclusion decision and extracted value against the primary sources. The authors take full responsibility for the content of this paper.

\bibliographystyle{IEEEtran}
\bibliography{references}

@inproceedings{Cao2021FLTrust,
  author    = {Cao, Xiaoyu and Fang, Minghong and Liu, Jia and Gong, Neil Zhenqiang},
  title     = {{FLTrust}: Byzantine-robust Federated Learning via Trust Bootstrapping},
  booktitle = {Proceedings of the 2021 Network and Distributed System Security Symposium},
  year      = {2021},
  doi       = {10.14722/ndss.2021.24434} 
}

@inproceedings{Fung2020FoolsGold,
  author    = {Fung, Clement and Yoon, Chris J. M. and Beschastnikh, Ivan},
  title     = {The Limitations of Federated Learning in Sybil Settings},
  booktitle = {Proceedings of the 23rd International Symposium on Research in Attacks, Intrusions and Defenses},
  year      = {2020},
  pages     = {301--316},
  publisher = {USENIX Association},
  url       = {https://www.usenix.org/conference/raid2020/presentation/fung}
}

@misc{Lin2023FLD,
  author    = {Lin, Yihang and Zhou, Pengyuan and Wu, Zhiqian and Liao, Yong},
  title     = {Mitigating Backdoors in Federated Learning with {FLD}},
  year      = {2023},
  eprint    = {2303.00302},
  archivePrefix = {arXiv},
  primaryClass = {cs.CR},
  url       = {https://arxiv.org/abs/2303.00302}
}

@inproceedings{Nguyen2022FLAME,
  author    = {Nguyen, Thien Duc and Rieger, Phillip and Chen, Huili and Yalame, Hossein and M{\"o}llering, Helen and Fereidooni, Hossein and Marchal, Samuel and Miettinen, Markus and Mirhoseini, Azalia and Zeitouni, Shaza and Koushanfar, Farinaz and Sadeghi, Ahmad-Reza and Schneider, Thomas},
  title     = {{FLAME}: Taming Backdoors in Federated Learning},
  booktitle = {31st USENIX Security Symposium},
  year      = {2022},
  pages     = {1415--1432},
  publisher = {USENIX Association},
  address   = {Boston, MA},
  url       = {https://www.usenix.org/conference/usenixsecurity22/presentation/nguyen}
}

@inproceedings{Rieger2024CrowdGuard,
  author    = {Rieger, Phillip and Krau{ss}, Torsten and Miettinen, Markus and Dmitrienko, Alexandra and Sadeghi, Ahmad-Reza},
  title     = {{CrowdGuard}: Federated Backdoor Detection in Federated Learning},
  booktitle = {Proceedings of the 2024 Network and Distributed System Security Symposium},
  year      = {2024},
  doi       = {10.14722/ndss.2024.23233} 
}

@inproceedings{Xie2021CRFL,
  author    = {Xie, Chulin and Chen, Minghao and Chen, Pin-Yu and Li, Bo},
  title     = {{CRFL}: Certifiably Robust Federated Learning against Backdoor Attacks},
  booktitle = {Proceedings of the 38th International Conference on Machine Learning},
  series    = {Proceedings of Machine Learning Research},
  volume    = {139},
  pages     = {11372--11382},
  year      = {2021},
  publisher = {PMLR},
  url       = {https://proceedings.mlr.press/v139/xie21a.html}
}

@article{Xu2024FedGNNBackdoors,
  author  = {Xu, Jing and Koffas, Stefanos and Picek, Stjepan},
  title   = {Unveiling the Threat: Investigating Distributed and Centralized Backdoor Attacks in Federated Graph Neural Networks},
  journal = {Digital Threats: Research and Practice},
  volume  = {5},
  number  = {2},
  articleno = {15},
  year    = {2024},
  doi     = {10.1145/3633206} 
}

@inproceedings{bagdasaryan2020,
 author={Eugene Bagdasaryan and Andreas Veit and Yiqing Hua and Deborah Estrin and Vitaly Shmatikov},
 title={How To Backdoor Federated Learning},
 booktitle={Proceedings of the 23rd International Conference on Artificial Intelligence and Statistics},
 series={Proceedings of Machine Learning Research},volume={108},pages={2938--2948},year={2020},
 url={https://proceedings.mlr.press/v108/bagdasaryan20a.html}}

@article{bellachia2026sok,
 author={Ahmed Ayoub Bellachia and Mouhamed Amine Bouchiha and Yacine Ghamri-Doudane},
 title={{SoK}: Understanding backdoor attacks \& defenses in federated learning},
 journal={Expert Systems with Applications}, volume={325}, pages={132558}, year={2026},
 doi={10.1016/j.eswa.2026.132558}}

@inproceedings{bonawitz2017,
 author={Keith Bonawitz and Vladimir Ivanov and Ben Kreuter and Antonio Marcedone and H. Brendan McMahan and Sarvar Patel and Daniel Ramage and Aaron Segal and Karn Seth},
 title={Practical Secure Aggregation for Privacy-Preserving Machine Learning},
 booktitle={Proceedings of the 2017 ACM SIGSAC Conference on Computer and Communications Security},
 year={2017},doi={10.1145/3133956.3133982} }

@article{djemaa2026survey,
 author={Aimen Djemaa and Djamel Djenouri and Phil Legg},
 title={Heterogeneity-Aware Poisoning Attacks and Mitigation in Federated Learning: A Comprehensive Survey and Taxonomy},
 journal={Electronics}, volume={15}, number={13}, pages={2876}, year={2026},
 doi={10.3390/electronics15132876}}

@inproceedings{feng2025sadba,
  title     = {{SADBA}: Self-Adaptive Distributed Backdoor Attack Against Federated Learning},
  author    = {Feng, Jun and Lai, Yuzhe and Sun, Hong and Ren, Bocheng},
  booktitle = {Proceedings of the AAAI Conference on Artificial Intelligence},
  volume    = {39},
  number    = {16},
  pages     = {16568--16576},
  year      = {2025},
  doi       = {10.1609/aaai.v39i16.33820}
}

@article{kim2025robust,
 author={Seunghan Kim and Changhoon Lim and Gwonsang Ryu and Hyunil Kim},
 title={How Robust Are Language Models against Backdoors in Federated Learning?},
 journal={Computer Modeling in Engineering \& Sciences},volume={145},number={2},pages={2617--2630},year={2025},doi={10.32604/cmes.2025.071190}}

@article{li2025survey,
 author={Zhaozheng Li and Jiahe Lan and Zheng Yan and Erol Gelenbe},
 title={Backdoor attacks and defense mechanisms in federated learning: A survey},
 journal={Information Fusion}, volume={123}, pages={103248}, year={2025},
 doi={10.1016/j.inffus.2025.103248}}

@misc{liu2024dmba,
 author={Tao Liu and Dapeng Man and Jiguang Lv and Chen Xu and Weiye Xi and Huanran Wang and Yuhang Zhang and Tianming Zhao and Wu Yang},
 title={Act in Collusion: Distributed Multi-Target Backdoor Attacks in Federated Learning},
 year={2026},howpublished={arXiv:2411.03926v3},
 url={https://arxiv.org/abs/2411.03926v3}}

@inproceedings{liu2024fcba,
  title     = {Beyond Traditional Threats: A Persistent Backdoor Attack on Federated Learning},
  author    = {Liu, Tao and Zhang, Yuhang and Feng, Zhu and Yang, Zhiqin and Xu, Chen and Man, Dapeng and Yang, Wu},
  booktitle = {Proceedings of the AAAI Conference on Artificial Intelligence},
  volume    = {38},
  number    = {19},
  pages     = {21359--21367},
  year      = {2024},
  doi       = {10.1609/aaai.v38i19.30131}
}

@inproceedings{liu2026dbadfl,
  title     = {{DBA-DFL}: Towards Distributed Backdoor Attacks with Network Detection in Decentralized Federated Learning},
  author    = {Liu, Bohan and Xiao, Yang and Ye, Ruimeng and Ling, Zinan and Ma, Xiaolong and Hui, Bo},
  booktitle = {Proceedings of the European Conference on Artificial Intelligence},
  year      = {2025},
  doi       = {10.3233/FAIA250961} 
}

@article{lyu2025coba,
  title   = {{CoBA}: Collusive Backdoor Attacks With Optimized Trigger to Federated Learning},
  author  = {Lyu, Xiaoting and Han, Yufei and Wang, Wei and Liu, Jingkai and Wang, Bin and Chen, Kai and Li, Yidong and Liu, Jiqiang and Zhang, Xiangliang},
  journal = {IEEE Transactions on Dependable and Secure Computing},
  volume  = {22},
  number  = {2},
  pages   = {1506--1518},
  year    = {2025},
  doi     = {10.1109/TDSC.2024.3445637}
}

@inproceedings{mcmahan2017,
 author={H. Brendan McMahan and Eider Moore and Daniel Ramage and Seth Hampson and Blaise {Ag{\"u}era y Arcas}},
 title={Communication-Efficient Learning of Deep Networks from Decentralized Data},
 booktitle={Proceedings of the 20th International Conference on Artificial Intelligence and Statistics},
 series={Proceedings of Machine Learning Research},volume={54},pages={1273--1282},year={2017},
 url={https://proceedings.mlr.press/v54/mcmahan17a.html}}

@article{ren2024sba,
  title   = {Shadow Backdoor Attack: Multi-intensity Backdoor Attack against Federated Learning},
  author  = {Ren, Qixian and Zheng, Yu and Yang, Chao and Li, Yue and Ma, Jianfeng},
  journal = {Computers \& Security},
  volume  = {139},
  pages   = {103740},
  year    = {2024},
  doi     = {10.1016/j.cose.2024.103740}
}

@article{wan2024survey,
 author={Yichen Wan and Youyang Qu and Wei Ni and Yong Xiang and Longxiang Gao and Ekram Hossain},
 title={Data and Model Poisoning Backdoor Attacks on Wireless Federated Learning, and the Defense Mechanisms: A Comprehensive Survey},
 journal={IEEE Communications Surveys \& Tutorials},volume={26},number={3},pages={1861--1897},year={2024},doi={10.1109/COMST.2024.3361451}}

@inproceedings{xie2020dba,
  title     = {{DBA}: Distributed Backdoor Attacks against Federated Learning},
  author    = {Xie, Chulin and Huang, Keli and Chen, Pin-Yu and Li, Bo},
  booktitle = {International Conference on Learning Representations},
  year      = {2020},
  url       = {https://openreview.net/forum?id=rkgyS0VFvr}
}

@inproceedings{yang2024optgdba,
  title     = {Distributed Backdoor Attacks on Federated Graph Learning and Certified Defenses},
  author    = {Yang, Yuxin and Li, Qiang and Jia, Jinyuan and Hong, Yuan and Wang, Binghui},
  booktitle = {Proceedings of the 2024 ACM SIGSAC Conference on Computer and Communications Security},
  pages     = {2829--2843},
  year      = {2024},
  doi       = {10.1145/3658644.3690187}
}

@inproceedings{jiang2026boba,
 author={Zhengyuan Jiang and Xingyu Lyu and Shanghao Shi and Yang Xiao and Yimin Chen and Y. Thomas Hou and Wenjing Lou and Ning Wang},
 title={{BoBa}: Boosting Backdoor Detection through Data Distribution Inference in Federated Learning},
 booktitle={ECAI 2025: 28th European Conference on Artificial Intelligence},
 year={2025},
 doi={10.3233/FAIA250914} 
}

@misc{zhao2026signal,
 author={Haodong Zhao and Jinming Hu and Gongshen Liu},
 title={Revisiting Backdoor Threat in Federated Instruction Tuning from a Signal Aggregation Perspective},
 year={2026},
 howpublished={arXiv:2602.15671v1},
 url={https://arxiv.org/abs/2602.15671v1}
}

@article{wang2025improved,
  author  = {Wang, Jian and Shen, Hong and Ke, Wei and Liu, Xue Hua},
  title   = {Improved Distributed Backdoor Attacks in Federated Learning by Density-Adaptive Data Poisoning and Projection-Based Gradient Updating},
  journal = {IEEE Access},
  volume  = {13},
  pages   = {130651--130671},
  year    = {2025},
  doi     = {10.1109/ACCESS.2025.3586416}
}

@misc{wang2025ftdba,
  author       = {Wang, Jian and Shen, Hong and Lam, Chan-Tong},
  title        = {Unveiling Hidden Threats: Hierarchical Self-Similar Fractal Triggers for Stealthy Distributed Backdoor Attacks in Federated Learning},
  year         = {2025},
  howpublished = {arXiv:2511.09252v2},
  url          = {https://arxiv.org/abs/2511.09252v2}
}

@misc{wang2026fdba,
  author       = {Wang, Jian and Shen, Hong and Ke, Wei and Liu, Xue Hua},
  title        = {Fine-grained Distributed Backdoor Attacks in Federated Learning},
  year         = {2026},
  howpublished = {arXiv:2609.07147v1},
  url          = {https://arxiv.org/abs/2609.07147}
}

@article{yang2026pba,
  author  = {Yang, Wenyin and Wu, Weidong and Liu, Fen and Ma, Li and Zhang, Zhiliang and Yang, Baoyao},
  title   = {{PBA}: Persistent Backdoor Attack on Federated Learning via Distributed Generative Triggers},
  journal = {Computer Networks},
  volume  = {277},
  pages   = {112038},
  year    = {2026},
  doi     = {10.1016/j.comnet.2026.112038}
}

@inproceedings{sundar2022swift,
 author={Agnideven Palanisamy Sundar and Feng Li and Xukai Zou and Tianchong Gao},
 title={Distributed Swift and Stealthy Backdoor Attack on Federated Learning},
 booktitle={2022 IEEE International Conference on Networking, Architecture and Storage (NAS)},
 year={2022},
 pages={1--8},
 url={https://scholarworks.indianapolis.iu.edu/items/97d8b673-e4f2-4a87-b621-62dca362e0b5}
}

@inproceedings{wang2024dynamic,
 author={Jian Wang and Hong Shen and Xuehua Liu and Hua Zhou and Yuli Li},
 title={Distributed Backdoor Attacks in Federated Learning Generated by {DynamicTriggers}},
 booktitle={Information Security Theory and Practice (WISTP 2024)},
 series={Lecture Notes in Computer Science},
 volume={14625},
 pages={178--193},
 year={2024},
 doi={10.1007/978-3-031-60391-4_12} 
}

@misc{hao2025mtfba,
 author={Lingguag Hao and Kuangrong Hao and Bing Wei and {Xue-song} Tang},
 title={Multi-Target Federated Backdoor Attack Based on Feature Aggregation},
 howpublished={arXiv:2502.16545v1},
 year={2025},
 url={https://arxiv.org/abs/2502.16545v1}
}

@inproceedings{Liu2020ClientEdgeCloudHierFL,
  author    = {Liu, Lumin and Zhang, Jun and Song, S. H. and Letaief, Khaled B.},
  title     = {Client-Edge-Cloud Hierarchical Federated Learning},
  booktitle = {2020 IEEE International Conference on Communications (ICC)},
  pages     = {1--6},
  year      = {2020},
  doi       = {10.1109/ICC40277.2020.9148862} 
}

@inproceedings{collapois2025,
 author={Phung Lai and Guanxiong Liu and NhatHai Phan and Issa Khalil and Abdallah Khreishah and Xintao Wu},
 title={A Client-level Assessment of Collaborative Backdoor Poisoning in Non-IID Federated Learning},
 booktitle={2025 IEEE 45th International Conference on Distributed Computing Systems (ICDCS)},
 pages={1--11}, year={2025}, doi={10.1109/ICDCS63083.2025.00010} 
}

@misc{wang2026structure,
 author={Jian Wang and Hong Shen and Wei Ke and Xue Hua Liu},
 title={Structure-Aware Distributed Backdoor Attacks in Federated Learning},
 year={2026}, eprint={2603.03865}, archivePrefix={arXiv},
 url={https://arxiv.org/abs/2603.03865}
}

@article{alharbi2024cba,
 author={Saier Alharbi and Yifan Guo and Wei Yu},
 title={Collusive Backdoor Attacks in Federated Learning Frameworks for {IoT} Systems},
 journal={IEEE Internet of Things Journal}, volume={11}, number={11}, pages={19694--19707}, year={2024}, doi={10.1109/JIOT.2024.3368754}
}

@inproceedings{lyu2023cerp,
 author={Xiaoting Lyu and Yufei Han and Wei Wang and Jingkai Liu and Bin Wang and Jiqiang Liu and Xiangliang Zhang},
 title={Poisoning with Cerberus: Stealthy and Colluded Backdoor Attack against Federated Learning},
 booktitle={Proceedings of the AAAI Conference on Artificial Intelligence},
 volume={37}, number={7}, pages={9020--9028}, year={2023}, doi={10.1609/aaai.v37i7.26083} 
}

@inproceedings{li2025nigdba,
 author={Ken Li and Bin Shi and Jiazhe Wei and Bo Dong},
 title={{NI-GDBA}: Non-Intrusive Distributed Backdoor Attack Based on Adaptive Perturbation on Federated Graph Learning},
 booktitle={Proceedings of the ACM Web Conference 2025}, pages={852--862},
 year={2025}, doi={10.1145/3696410.3714630} 
}

@misc{abad2024timebandits,
 author={Gorka Abad and Stjepan Picek and Aitor Urbieta},
 title={Time-Distributed Backdoor Attacks on Federated Spiking Learning},
 year={2024}, howpublished={arXiv:2402.02886v1},
 url={https://arxiv.org/abs/2402.02886v1}
}

@inproceedings{yang2022cooperative,
 author={Jijia Yang and Jiangang Shu and Xiaohua Jia},
 title={Breaking Distributed Backdoor Defenses for Federated Learning in Non-{IID} Settings},
 booktitle={2022 18th International Conference on Mobility, Sensing and Networking (MSN)},
 year={2022},pages={347--354}, doi={10.1109/MSN57253.2022.00064} 
}

@article{yang2025ssrdba,
 author={Yuxin Yang and Qiang Li and Yuede Ji and Binghui Wang},
 title={A Secret Sharing-Inspired Robust Distributed Backdoor Attack to Federated Learning},
 journal={ACM Transactions on Privacy and Security},volume={28},number={2},
 pages={27:1--27:19},year={2025},doi={10.1145/3725814} 
}

@inproceedings{ai2025sata,
 author={Xin Ai and Yang Cao},
 title={{SATA}: A Distributed Stealthy and Persistent Backdoor Attack via Trigger Association in Federated Learning},
 booktitle={2025 IEEE 24th International Conference on Trust, Security and Privacy in Computing and Communications (TrustCom)},
 year={2025},pages={2598--2605},doi={10.1109/Trustcom66490.2025.00306} 
}

@inproceedings{blanchard2017krum,
 author={Peva Blanchard and El Mahdi El Mhamdi and Rachid Guerraoui and Julien Stainer},
 title={Machine Learning with Adversaries: {Byzantine} Tolerant Gradient Descent},
 booktitle={Advances in Neural Information Processing Systems},volume={30},year={2017}}

@inproceedings{yin2018trimmed,
 author={Dong Yin and Yudong Chen and Kannan Ramchandran and Peter Bartlett},
 title={{Byzantine}-Robust Distributed Learning: Towards Optimal Statistical Rates},
 booktitle={Proceedings of the 35th International Conference on Machine Learning},
 series={Proceedings of Machine Learning Research},volume={80},pages={5650--5659},year={2018}}

@article{pillutla2022rfa,
 author={Krishna Pillutla and Sham M. Kakade and Zaid Harchaoui},
 title={Robust Aggregation for Federated Learning},
 journal={IEEE Transactions on Signal Processing},volume={70},pages={1142--1154},year={2022},
 doi={10.1109/TSP.2022.3153135}}

@inproceedings{ozdayi2021rlr,
 author={Mustafa Safa Ozdayi and Murat Kantarcioglu and Yulia R. Gel},
 title={Defending against Backdoors in Federated Learning with Robust Learning Rate},
 booktitle={Proceedings of the AAAI Conference on Artificial Intelligence},volume={35},number={10},pages={9268--9276},year={2021}}

@misc{sun2019normbound,
 author={Ziteng Sun and Peter Kairouz and Ananda Theertha Suresh and H. Brendan McMahan},
 title={Can You Really Backdoor Federated Learning?},
 year={2019},howpublished={arXiv:1911.07963},url={https://arxiv.org/abs/1911.07963}}

@inproceedings{mcmahan2018dpfedavg,
 author={H. Brendan McMahan and Daniel Ramage and Kunal Talwar and Li Zhang},
 title={Learning Differentially Private Recurrent Language Models},
 booktitle={International Conference on Learning Representations},year={2018}}

@inproceedings{rieger2022deepsight,
 author={Phillip Rieger and Thien Duc Nguyen and Markus Miettinen and Ahmad-Reza Sadeghi},
 title={{DeepSight}: Mitigating Backdoor Attacks in Federated Learning Through Deep Model Inspection},
 booktitle={Proceedings of the 2022 Network and Distributed System Security Symposium},year={2022},
 doi={10.14722/ndss.2022.23156}}

@inproceedings{zhang2023flip,
 author={Kaiyuan Zhang and Guanhong Tao and Qiuling Xu and Siyuan Cheng and Shengwei An and Yingqi Liu and Shiwei Feng and Guangyu Shen and Pin-Yu Chen and Shiqing Ma and Xiangyu Zhang},
 title={{FLIP}: A Provable Defense Framework for Backdoor Mitigation in Federated Learning},
 booktitle={International Conference on Learning Representations},year={2023}}

@inproceedings{jia2023fedgame,
 author={Jinyuan Jia and Zhuowen Yuan and Dinuka Sahabandu and Luyao Niu and Arezoo Rajabi and Bhaskar Ramasubramanian and Bo Li and Radha Poovendran},
 title={{FedGame}: A Game-Theoretic Defense against Backdoor Attacks in Federated Learning},
 booktitle={Advances in Neural Information Processing Systems},volume={36},year={2023}}

@inproceedings{fereidooni2024freqfed,
 author={Hossein Fereidooni and Alessandro Pegoraro and Phillip Rieger and Alexandra Dmitrienko and Ahmad-Reza Sadeghi},
 title={{FreqFed}: A Frequency Analysis-Based Approach for Mitigating Poisoning Attacks in Federated Learning},
 booktitle={Proceedings of the 2024 Network and Distributed System Security Symposium},year={2024}}

@inproceedings{li2024backdoorindicator,
 author={Songze Li and Yanbo Dai},
 title={{BackdoorIndicator}: Leveraging {OOD} Data for Proactive Backdoor Detection in Federated Learning},
 booktitle={33rd USENIX Security Symposium},year={2024}}

@inproceedings{cao2023fedrecover,
 author={Xiaoyu Cao and Jinyuan Jia and Zaixi Zhang and Neil Zhenqiang Gong},
 title={{FedRecover}: Recovering from Poisoning Attacks in Federated Learning using Historical Information},
 booktitle={2023 IEEE Symposium on Security and Privacy (SP)},year={2023}}

@inproceedings{xu2022signguard,
 author={Jian Xu and Shao-Lun Huang and Linqi Song and Tian Lan},
 title={{Byzantine}-robust Federated Learning through Collaborative Malicious Gradient Filtering},
 booktitle={2022 IEEE 42nd International Conference on Distributed Computing Systems (ICDCS)},year={2022}}

@misc{li2023clippedclustering,
 author={Shenghui Li and Edith Ngai and Thiemo Voigt},
 title={An Experimental Study of {Byzantine}-Robust Aggregation Schemes in Federated Learning},
 year={2023},howpublished={arXiv:2302.07173},url={https://arxiv.org/abs/2302.07173}}

@inproceedings{wang2019neuralcleanse,
 author={Bolun Wang and Yuanshun Yao and Shawn Shan and Huiying Li and Bimal Viswanath and Haitao Zheng and Ben Y. Zhao},
 title={Neural Cleanse: Identifying and Mitigating Backdoor Attacks in Neural Networks},
 booktitle={2019 IEEE Symposium on Security and Privacy (SP)},pages={707--723},year={2019}}

@misc{chen2018activation,
 author={Bryant Chen and Wilka Carvalho and Nathalie Baracaldo and Heiko Ludwig and Benjamin Edwards and Taesung Lee and Ian Molloy and Biplav Srivastava},
 title={Detecting Backdoor Attacks on Deep Neural Networks by Activation Clustering},
 year={2018},howpublished={arXiv:1811.03728},url={https://arxiv.org/abs/1811.03728}}

@inproceedings{gao2019strip,
 author={Yansong Gao and Change Xu and Derui Wang and Shiping Chen and Damith C. Ranasinghe and Surya Nepal},
 title={{STRIP}: A Defence against Trojan Attacks on Deep Neural Networks},
 booktitle={Proceedings of the 35th Annual Computer Security Applications Conference},pages={113--125},year={2019}}

@inproceedings{liu2018fineprune,
 author={Kang Liu and Brendan Dolan-Gavitt and Siddharth Garg},
 title={Fine-Pruning: Defending against Backdooring Attacks on Deep Neural Networks},
 booktitle={Research in Attacks, Intrusions, and Defenses (RAID)},series={Lecture Notes in Computer Science},volume={11050},pages={273--294},year={2018}}

@misc{wu2020fedprune,
 author={Chen Wu and Xian Yang and Sencun Zhu and Prasenjit Mitra},
 title={Mitigating Backdoor Attacks in Federated Learning},
 year={2020},howpublished={arXiv:2011.01767},url={https://arxiv.org/abs/2011.01767}}

@inproceedings{li2020fedprox,
 author={Tian Li and Anit Kumar Sahu and Manzil Zaheer and Maziar Sanjabi and Ameet Talwalkar and Virginia Smith},
 title={Federated Optimization in Heterogeneous Networks},
 booktitle={Proceedings of Machine Learning and Systems},volume={2},pages={429--450},year={2020}}

@inproceedings{wang2020fednova,
 author={Jianyu Wang and Qinghua Liu and Hao Liang and Gauri Joshi and H. Vincent Poor},
 title={Tackling the Objective Inconsistency Problem in Heterogeneous Federated Optimization},
 booktitle={Advances in Neural Information Processing Systems},volume={33},year={2020}}

@inproceedings{lian2017dpsgd,
 author={Xiangru Lian and Ce Zhang and Huan Zhang and Cho-Jui Hsieh and Wei Zhang and Ji Liu},
 title={Can Decentralized Algorithms Outperform Centralized Algorithms? {A} Case Study for Decentralized Parallel Stochastic Gradient Descent},
 booktitle={Advances in Neural Information Processing Systems},volume={30},year={2017}}

@inproceedings{bornstein2023swift,
 author={Marco Bornstein and Tahseen Rabbani and Evan Wang and Amrit Singh Bedi and Furong Huang},
 title={{SWIFT}: Rapid Decentralized Federated Learning via Wait-Free Model Communication},
 booktitle={International Conference on Learning Representations},year={2023}}

@inproceedings{bhagoji2019lens,
 author={Arjun Nitin Bhagoji and Supriyo Chakraborty and Prateek Mittal and Seraphin Calo},
 title={Analyzing Federated Learning through an Adversarial Lens},
 booktitle={Proceedings of the 36th International Conference on Machine Learning},
 series={Proceedings of Machine Learning Research},volume={97},pages={634--643},year={2019}}

@inproceedings{wang2020tails,
 author={Hongyi Wang and Kartik Sreenivasan and Shashank Rajput and Harit Vishwakarma and Saurabh Agarwal and {Jy-yong} Sohn and Kangwook Lee and Dimitris Papailiopoulos},
 title={Attack of the Tails: Yes, You Really Can Backdoor Federated Learning},
 booktitle={Advances in Neural Information Processing Systems},volume={33},year={2020}}

@inproceedings{zhang2022neurotoxin,
 author={Zhengming Zhang and Ashwinee Panda and Linyue Song and Yaoqing Yang and Michael Mahoney and Prateek Mittal and Kannan Ramchandran and Joseph Gonzalez},
 title={Neurotoxin: Durable Backdoors in Federated Learning},
 booktitle={Proceedings of the 39th International Conference on Machine Learning},
 series={Proceedings of Machine Learning Research},volume={162},pages={26429--26446},year={2022}}

@inproceedings{dai2023chameleon,
 author={Yanbo Dai and Songze Li},
 title={Chameleon: Adapting to Peer Images for Planting Durable Backdoors in Federated Learning},
 booktitle={Proceedings of the 40th International Conference on Machine Learning},
 series={Proceedings of Machine Learning Research},volume={202},year={2023}}

@inproceedings{fang2023f3ba,
 author={Pei Fang and Jinghui Chen},
 title={On the Vulnerability of Backdoor Defenses for Federated Learning},
 booktitle={Proceedings of the AAAI Conference on Artificial Intelligence},volume={37},number={10},year={2023}}

@inproceedings{zhang2023a3fl,
 author={Hangfan Zhang and Jinyuan Jia and Jinghui Chen and Lu Lin and Dinghao Wu},
 title={{A3FL}: Adversarially Adaptive Backdoor Attacks to Federated Learning},
 booktitle={Advances in Neural Information Processing Systems},volume={36},year={2023}}

@inproceedings{nguyen2023iba,
 author={Thuy Dung Nguyen and Tuan A. Nguyen and Anh Tran and Khoa D. Doan and Kok-Seng Wong},
 title={{IBA}: Towards Irreversible Backdoor Attacks in Federated Learning},
 booktitle={Advances in Neural Information Processing Systems},volume={36},year={2023}}

@inproceedings{li2023threedfed,
 author={Haoyang Li and Qingqing Ye and Haibo Hu and Jin Li and Leixia Wang and Chengfang Fang and Jie Shi},
 title={{3DFed}: Adaptive and Extensible Framework for Covert Backdoor Attack in Federated Learning},
 booktitle={2023 IEEE Symposium on Security and Privacy (SP)},year={2023}}

@inproceedings{li2024darkfed,
 author={Minghui Li and Wei Wan and Yuxuan Ning and Shengshan Hu and Lulu Xue and Leo Yu Zhang and Yichen Wang},
 title={{DarkFed}: A Data-Free Backdoor Attack in Federated Learning},
 booktitle={Proceedings of the 33rd International Joint Conference on Artificial Intelligence},year={2024}}

@inproceedings{naseri2024badvfl,
 author={Mohammad Naseri and Yufei Han and Emiliano {De Cristofaro}},
 title={{BadVFL}: Backdoor Attacks in Vertical Federated Learning},
 booktitle={2024 IEEE Symposium on Security and Privacy (SP)},year={2024}}

@misc{dao2025backfed,
 author={Thinh Dao and Dung Thuy Nguyen and Khoa D. Doan and Kok-Seng Wong},
 title={{BackFed}: A Standardized and Efficient Benchmark Framework for Evaluating Backdoor Attacks in Federated Learning},
 year={2025},howpublished={arXiv:2507.04903},url={https://arxiv.org/abs/2507.04903}}

@article{gong2022coordinated,
 author={Xueluan Gong and Yanjiao Chen and Huayang Huang and Yuqing Liao and Shuai Wang and Qian Wang},
 title={Coordinated Backdoor Attacks against Federated Learning with Model-Dependent Triggers},
 journal={IEEE Network},volume={36},number={1},pages={84--90},year={2022},doi={10.1109/MNET.011.2000783}}

@misc{syros2023p2p,
 author={Georgios Syros and Gokberk Yar and Simona Boboila and Cristina Nita-Rotaru and Alina Oprea},
 title={Backdoor Attacks in Peer-to-Peer Federated Learning},
 year={2023},howpublished={arXiv:2301.09732},url={https://arxiv.org/abs/2301.09732}}

@article{rodriguez2023threats,
 author={Nuria Rodr{\'i}guez-Barroso and Daniel Jim{\'e}nez-L{\'o}pez and M. Victoria Luz{\'o}n and Francisco Herrera and Eugenio Mart{\'i}nez-C{\'a}mara},
 title={Survey on Federated Learning Threats: Concepts, Taxonomy on Attacks and Defences, Experimental Study and Challenges},
 journal={Information Fusion},volume={90},pages={148--173},year={2023}}

@article{nguyen2024backdoorsurvey,
 author={Thuy Dung Nguyen and Tuan Nguyen and Phi Le Nguyen and Hieu H. Pham and Khoa D. Doan and Kok-Seng Wong},
 title={Backdoor Attacks and Defenses in Federated Learning: Survey, Challenges and Future Research Directions},
 journal={Engineering Applications of Artificial Intelligence},volume={127},pages={107166},year={2024}}

@inproceedings{zhang2022fldetector,
 author={Zaixi Zhang and Xiaoyu Cao and Jinyuan Jia and Neil Zhenqiang Gong},
 title={{FLDetector}: Defending Federated Learning Against Model Poisoning Attacks via Detecting Malicious Clients},
 booktitle={Proceedings of the 28th ACM SIGKDD Conference on Knowledge Discovery and Data Mining},year={2022}}

@inproceedings{krauss2023mesas,
 author={Torsten Krau{\ss} and Alexandra Dmitrienko},
 title={{MESAS}: Poisoning Defense for Federated Learning Resilient against Adaptive Attackers},
 booktitle={Proceedings of the 2023 ACM SIGSAC Conference on Computer and Communications Security},year={2023}}

@inproceedings{huang2023lockdown,
 author={Tiansheng Huang and Sihao Hu and Ka-Ho Chow and Fatih Ilhan and Selim Tekin and Ling Liu},
 title={Lockdown: Backdoor Defense for Federated Learning with Isolated Subspace Training},
 booktitle={Advances in Neural Information Processing Systems},volume={36},year={2023}}

@inproceedings{qin2024snowball,
 author={Zhen Qin and Feiyi Chen and Chen Zhi and Xueqiang Yan and Shuiguang Deng},
 title={Resisting Backdoor Attacks in Federated Learning via Bidirectional Elections and Individual Perspective},
 booktitle={Proceedings of the AAAI Conference on Artificial Intelligence},volume={38},pages={14677--14685},year={2024}}

@inproceedings{kabir2024flshield,
 author={Ehsanul Kabir and Zeyu Song and Md Rafi Ur Rashid and Shagufta Mehnaz},
 title={{FLShield}: A Validation Based Federated Learning Framework to Defend Against Poisoning Attacks},
 booktitle={IEEE Symposium on Security and Privacy},year={2024}}

@inproceedings{xu2025alignins,
 author={Jiahao Xu and Zikai Zhang and Rui Hu},
 title={Detecting Backdoor Attacks in Federated Learning via Direction Alignment Inspection},
 booktitle={Proceedings of the IEEE/CVF Conference on Computer Vision and Pattern Recognition},year={2025}}

@inproceedings{xie2024fedredefense,
 author={Yueqi Xie and Minghong Fang and Neil Zhenqiang Gong},
 title={{FedREDefense}: Defending against Model Poisoning Attacks for Federated Learning using Model Update Reconstruction Error},
 booktitle={Proceedings of the 41st International Conference on Machine Learning},series={PMLR},volume={235},pages={54460--54474},year={2024}}

@inproceedings{zhuang2024lp,
 author={Haomin Zhuang and Mingxian Yu and Hao Wang and Yang Hua and Jian Li and Xu Yuan},
 title={Backdoor Federated Learning by Poisoning Backdoor-Critical Layers},
 booktitle={International Conference on Learning Representations},year={2024}}

\appendices
\section{Data Availability}\label{app:availability}
The artifact is available at \url{https://anonymous.4open.science/r/fl-pieces-artifact-BFF3/}. It contains the experiment code with the unit tests of each experiment, the frozen configurations, the per-run and per-seed results for all 15 piece subsets, the per-round evaluations, the repair endpoints, the resource audit, and the written reports with all deviations and failures. It also contains the scripts that produce every figure in Section~\ref{sec:exp} and the coding of the downstream audit. Local paths in the code and configurations are replaced by placeholders. Model checkpoints and per-update statistics exceed the hosting limits, and we will release them with the final artifact. Upon acceptance, the artifact will be released under an open-source license.

\section{Review Method}\label{app:method}
\subsection{Search Design}
We combine a structured database search with citation- and method-led retrieval of primary studies, and we report them following the PRISMA 2020 flow structure. We ran the database search on 5--6 November 2026, with 5 November 2026 as the public-availability cutoff and no lower date bound. We searched IEEE Xplore, Scopus, and Web of Science Core Collection with the Boolean expression
\begin{quote}\small\raggedright
\texttt{federated AND ("distributed backdoor" OR "collusive backdoor" OR "collaborative backdoor" OR "split trigger")}
\end{quote}
over all fields (Xplore), title, abstract, and keywords (Scopus), and Topic (Web of Science), without language or document-type filters. The search returned 23, 48, and 15 records, respectively.

Because the exact phrases miss studies that describe coordination in other words, supplementary retrieval followed backward and forward citations of the core studies, named methods, author pages, arXiv, OpenReview, and publisher records. Supplementary queries also covered graph, decentralized, multi-target, IoT, and survey terminology. We identified single-client attacks and defenses that serve as comparators by the same citation route. They are cited as context and are not counted as primary studies.

\subsection{Eligibility and Screening}
An eligible attack study specifies malicious training contributions distributed across FL clients that implant a conditional targeted behavior in the shared model. Complementary contributions may be assigned in advance or coordinated during training. In the core group, the clients split trigger information among themselves. Extensions coordinate attacker roles, update alignment, target interference, or propagation over a decentralized topology. We retain test-time perturbations without a training-stage implantation step as adjacent evidence. We exclude studies without FL, conditional targeting, or an explicit coordination mechanism. Extraction requires an English full text, and we include preprints with their version identified. Surveys and defenses are retained as context rather than counted as attack studies.

Duplicates were removed by exact DOI and normalized-title matching. Conference and journal versions of the same work were merged into one study lineage when the journal version explicitly extends the conference version (the ACSAC 2022/DTRP 2024 graph-backdoor pair and the AAAI 2023/TDSC 2025 CerP/CoBA pair). Experimental conditions are attributed to the version that reports them. Title and abstract screening and full-text extraction used large language model (LLM) based assistants in three independent passes: an initial screen, a re-assessment of eligible candidates, and an audit of key sources. The authors checked every inclusion decision and every extracted value against the primary source. They also made the final decisions on context classification, version linkage, and threat-model boundaries. No independent human reviewer duplicated the first screening pass, which we report as a limitation below.
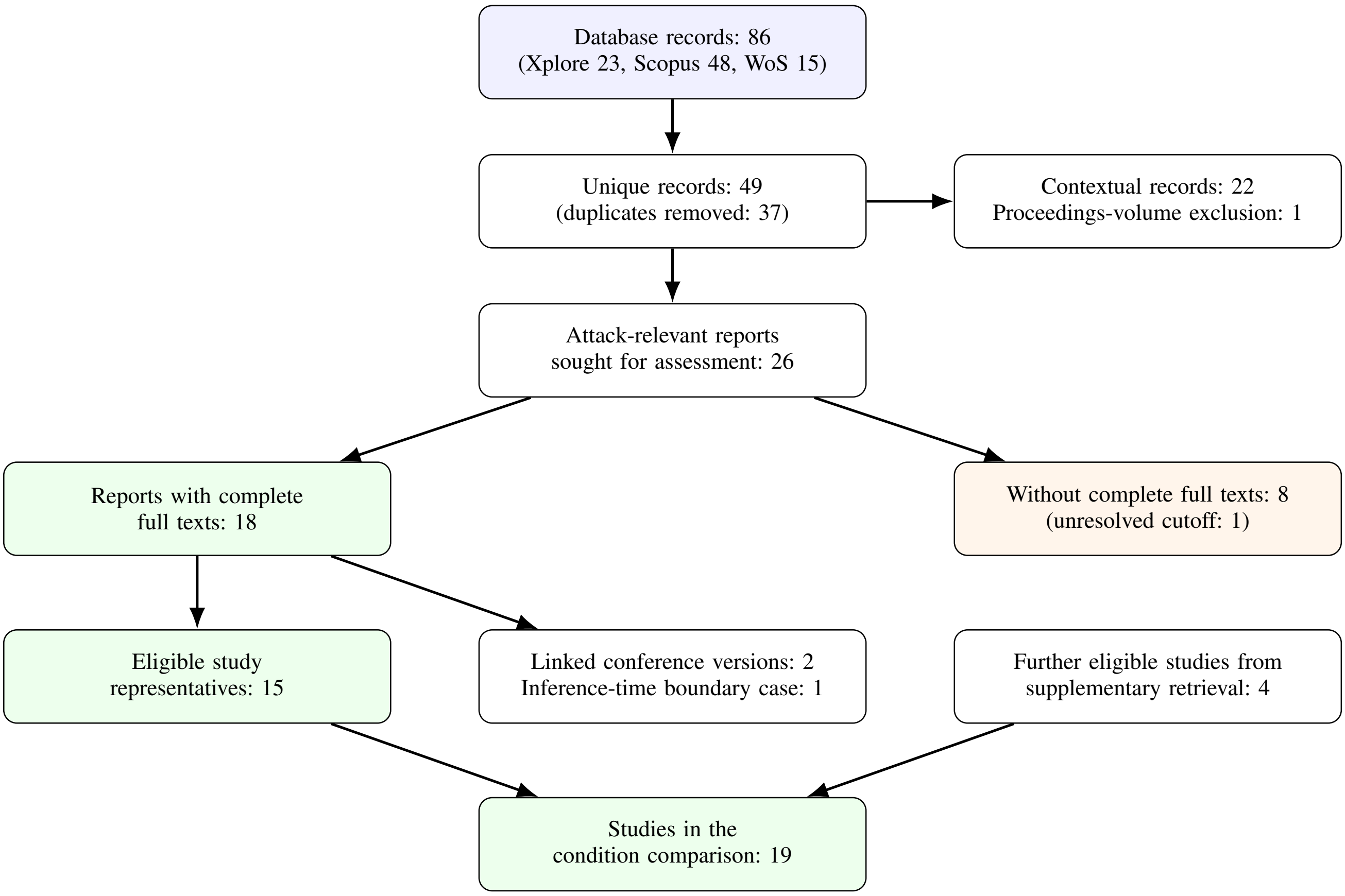
\begin{figure*}[t]
\centering
\resizebox{\linewidth}{!}{%
\begin{tikzpicture}[box/.style={draw,rounded corners,align=center,font=\scriptsize,text width=3.8cm,minimum height=1cm,inner sep=5pt},arr/.style={-{Latex},thick}]
\node[box,fill=blue!6] (raw) at (0,0) {Database records: 86\\(Xplore 23, Scopus 48, WoS 15)};
\node[box] (unique) at (0,-1.6) {Unique records: 49\\(duplicates removed: 37)};
\node[box] (context) at (5.1,-1.6) {Contextual records: 22\\Proceedings-volume exclusion: 1};
\node[box] (sought) at (0,-3.2) {Attack-relevant reports\\sought for assessment: 26};
\node[box,fill=green!7] (read) at (-5.1,-4.9) {Reports with complete\\full texts: 18};
\node[box,fill=orange!8] (await) at (5.1,-4.9) {Without complete full texts: 8\\(unresolved cutoff: 1)};
\node[box,fill=green!7] (eligible) at (-5.1,-6.7) {Eligible study\\representatives: 15};
\node[box] (linked) at (0,-6.7) {Linked conference versions: 2\\Inference-time boundary case: 1};
\node[box] (supp) at (5.1,-6.7) {Further eligible studies from\\supplementary retrieval: 4};
\node[box,fill=green!7] (final) at (0,-8.5) {Studies in the\\condition comparison: 19};
\draw[arr] (raw)--(unique);\draw[arr] (unique)--(context);\draw[arr] (unique)--(sought);\draw[arr] (sought)--(read);\draw[arr] (sought)--(await);\draw[arr] (read)--(eligible);\draw[arr] (read)--(linked);\draw[arr] (eligible)--(final);\draw[arr] (supp)--(final);
\end{tikzpicture}%
}
\caption{PRISMA-style record flow. Duplicate removal, version linkage, threat-boundary decisions, and unavailable full texts are separate operations. The four studies from supplementary retrieval are FCBA, FTDBA, FDBA, and TFI.}\label{fig:search-flow}
\end{figure*}

\subsection{Extraction and Synthesis}
For each study we extract the version used, malicious organization, topology, task and architecture, client population and participation, poisoning and update budgets, attack timing, post-attack continuation, tested defenses, and the section or table that reports each value. Values not stated in the source are recorded as not reported (NR) rather than inferred from related papers. Gong et al.~\cite{gong2022coordinated} and the peer-to-peer study of Syros et al.~\cite{syros2023p2p} were not retrieved by the structured search and are discussed but not coded. Table~\ref{tab:coding} summarizes these fields, and Appendix~\ref{app:extraction} gives the full extraction.

The synthesis compares mechanisms and experimental conditions and does not pool attack success rates. This is because the studies differ in ways that change the meaning of a success score. For example, the original DBA image setup forces malicious participation during multi-shot attacks, FCBA measures behavior after injection ends, and CollaPois averages outcomes over clients. Three comparisons organize the results: composition versus resources, exposure versus persistence, and observer versus protected unit.

\subsection{Limitations of the Review}
The structured search is reproducible but bounded by its phrase query. Supplementary retrieval reduces, but does not eliminate, missed studies. Eight attack-relevant reports could not be assessed in full text (Appendix~\ref{app:extraction}), and screening was not fully duplicated. We did not reproduce the attacks of individual studies and did not perform a meta-analysis. Instead, Section~\ref{sec:exp} tests the central comparison in a controlled experiment.

\subsection{Downstream Audit}\label{app:downstream}
The downstream audit asks how papers that \emph{use} DBA configure and measure it. Candidates were the defenses and attacks most often used as baselines in the coded studies and in BackFed~\cite{dao2025backfed}, supplemented by keyword searches for federated backdoor defenses and attacks that evaluate DBA. We included peer-reviewed papers at major security and machine-learning venues whose experiments run DBA, and coded them from the full text. The sample is purposive, not a census. Of 23 screened papers, 15 run DBA (Table~\ref{tab:downstream}). The other eight discuss DBA only as background or evaluate other attacks: CrowdGuard~\cite{Rieger2024CrowdGuard}, MESAS~\cite{krauss2023mesas}, FLIP~\cite{zhang2023flip}, CRFL~\cite{Xie2021CRFL}, DarkFed~\cite{li2024darkfed}, BayBFed (IEEE S\&P 2023), PFedBA (USENIX Security 2024), and a robustness study of personalized FL (KDD 2023). A complete-trigger attack counts as \emph{inferable} when it runs with the same attacker count as DBA in the same setup but is not presented as a comparison with it.

\begin{table*}[!htbp]
\caption{Downstream audit: how fifteen peer-reviewed papers configure and measure DBA. Role: D, defense evaluated against DBA; A, attack using DBA as a baseline. NR: not reported.}\label{tab:downstream}
\centering\scriptsize\setlength{\tabcolsep}{4pt}
\begin{tabular}{llcllccc}
\toprule
Paper & Venue & Role & DBA attackers & Forced or sampled & Scaling & Equal-attacker complete trigger & Pieces tested\\
\midrule
RLR~\cite{ozdayi2021rlr} & AAAI 2021 & D & 4 of 40 & NR & NR & inferable & no\\
FLAME~\cite{Nguyen2022FLAME} & USENIX Sec.\ 2022 & D & NR & NR & NR & no & no\\
DeepSight~\cite{rieger2022deepsight} & NDSS 2022 & D & NR & NR & NR & no & no\\
F3BA~\cite{fang2023f3ba} & AAAI 2023 & A & 4 of 20 & sampled & NR & no & no\\
A3FL~\cite{zhang2023a3fl} & NeurIPS 2023 & A & 1--20 of 100 & sampled & NR & no & no\\
IBA~\cite{nguyen2023iba} & NeurIPS 2023 & A & 4 & forced, every 10 rounds & NR & no & no\\
Lockdown~\cite{huang2023lockdown} & NeurIPS 2023 & D & NR & NR & NR & no & no\\
FedGame~\cite{jia2023fedgame} & NeurIPS 2023 & D & 6 of 10 & all selected & NR & no & no\\
FreqFed~\cite{fereidooni2024freqfed} & NDSS 2024 & D & NR & NR & NR & no & no\\
Snowball~\cite{qin2024snowball} & AAAI 2024 & D & 20\% & NR & NR & inferable & no\\
FLShield~\cite{kabir2024flshield} & IEEE S\&P 2024 & D & 40\% & NR & NR & no & no\\
BackdoorIndicator~\cite{li2024backdoorindicator} & USENIX Sec.\ 2024 & D & 40\% & every round & NR & inferable & no\\
LP attack~\cite{zhuang2024lp} & ICLR 2024 & A & 1 per round & forced & NR & yes & no\\
FedREDefense~\cite{xie2024fedredefense} & ICML 2024 & D & 28 of 100 & all participate & 3 & no & no\\
AlignIns~\cite{xu2025alignins} & CVPR 2025 & D & 20\% & NR & NR & no & no\\
\bottomrule
\end{tabular}
\end{table*}

\section{Detailed Extraction}\label{app:extraction}
Tables~\ref{tab:primary-extraction}--\ref{tab:new-fulltexts} report the conditions behind the coding in Table~\ref{tab:coding}, with the section or table of each source that supports them. Source-internal inconsistencies are retained rather than resolved.

\begin{table*}[!htbp]
\caption{Participation, task, and poisoning conditions of nine primary studies. NR: not reported in the source version used.}\label{tab:primary-extraction}
\centering\scriptsize\setlength{\tabcolsep}{3pt}
\begin{tabularx}{\textwidth}{P{0.14\textwidth}P{0.18\textwidth}P{0.25\textwidth}YY}
\toprule
Study / version & System and task & Participation & Poison/update budget and schedule & Source anchors\\
\midrule
DBA~\cite{xie2020dba} (ICLR 2020) & Server FL; MNIST, CIFAR-10, Tiny-ImageNet, LOAN & Images: 100 clients, 10 selected, 4 attackers. LOAN: 51 parties, 3 attackers. A-M forces attacker participation. & Poison ratios: LOAN 10/64, MNIST 20/64, CIFAR 5/64, Tiny 20/64. Replacement $\gamma=100$ images, 30 LOAN. A-S MNIST rounds 12/14/16/18. & Table 1; \S3.2; \S4.1; App.~A.1.\\
FCBA~\cite{liu2024fcba} (AAAI 2024) & Server FL; MNIST, CIFAR-10, GTSRB & 100 total, 10 selected/round; 14 malicious in default; malicious selected count/round NR. One update per malicious participant. & Single-shot; poison ratios 3/64, 2/64, 4/64 by task; $\gamma=100$ in headline comparison; CIFAR example completes at round 340. & pp. 2--6, Tables 1--3, Figs. 1--11.\\
SADBA~\cite{feng2025sadba} (AAAI 2025) & Server FL; MNIST, Fashion-MNIST, CIFAR-10 & Small/large settings 100/500; training text says select 10 from 100; minimum PMC is population threshold, not per-round attackers. & Single-shot; adaptive local poison/model optimization; one default poison rate NR. & pp. 2--6, Tables 1--3, Figs. 2,5--6.\\
DMBA~\cite{liu2024dmba} (arXiv v3) & Server FedAvg; CIFAR-10/100, GTSRB & Three malicious clients default; selected population/round NR. & First attack round 202; three attack rounds; replay ratio sweep; absolute poisoned count NR. & \S\S{}III--V; Tables III, XI, XIII--XIV; Fig. 7.\\
DBA-DFL~\cite{liu2026dbadfl} (ECAI 2025) & Peer-to-peer; CIFAR-10, MNIST; ring/grid-family & 40 clients, 10 attackers, 3 groups; attacker placement varied. & Total poisoned pixels equalized across methods; per-client poison ratio NR; distance inference first 100 epochs, 20 repeats. & \S\S{}3.1--3.2, 5; Tables 1--3, Figs. 2--9.\\
FTDBA~\cite{wang2025ftdba} (arXiv v2) & Server FedAvg; CIFAR-10, Tiny-ImageNet, GTSRB, ImageNet-100 & 100 clients; pool of 10 controlled; active attackers 2/4/6/8/10 tested. & Poison rate 30\%; 200 rounds CIFAR/GTSRB, 300 others; 5 local epochs. & \S{}VI; Tables I--IV, VIII--XIV.\\
FDBA~\cite{wang2026fdba} (arXiv v1) & Server FL; CIFAR-10 main setup; broader model/task table incompletely extracted & 10\% malicious share; per-round sampling NR. & Poison ratio sweep 1--10\%, includes 4\%; absolute poisoned count and total rounds NR. & \S\S{}3--7, App. A; Tables 3,5,8.\\
CollaPois~\cite{collapois2025} (arXiv v2) & Non-IID FL; FEMNIST and Sentiment140 & 3,400/5,600 clients by task; 1\% random participation; controlled shares 0.1/0.5/1\% studied. & Shared Trojaned model; aligned gradients; Dirichlet $\alpha=0.01$--100; target label 0. & \S\S{}III--V, App. A/E/G.\\
TFI~\cite{wang2026structure} (arXiv v1) & Simulated FL; CIFAR-10/ImageNet-100, ResNet-18, DenseNet-121, VGG-16, ViT-Base & 100 clients; 10\% randomly selected/round; attacker share NR. & Non-IID Dirichlet $\alpha=0.5$; 5 local epochs; 10\% default poison; 5\% ablation. & \S\S{}4.3, 6.1--6.7; Tables 1--3.\\
\bottomrule
\end{tabularx}
\end{table*}

\begin{table*}[!htbp]
\caption{Persistence and defense evidence for the same primary studies. Metrics and post-attack intervals are source-specific and are not standardized across papers.}\label{tab:study-outcomes}
\centering\scriptsize\setlength{\tabcolsep}{3pt}
\begin{tabularx}{\textwidth}{P{0.13\textwidth}P{0.27\textwidth}P{0.28\textwidth}Y}
\toprule
Study & Persistence or continuation & Defenses actually evaluated or claimed & Evidence boundary\\
\midrule
DBA~\cite{xie2020dba} & Single-shot and multi-shot schedules distinguished; benign continuation after staged injection. & RFA~\cite{pillutla2022rfa}/FoolsGold under A-M; appendix also describes Byzantine aggregation tests. & \S3.2, App.~A.5.\\
FCBA~\cite{liu2024fcba} & ASR-t at 0/40/80/120 benign rounds; at 120: 99.52/79.52/57.78\% on MNIST/CIFAR/GTSRB. & Participant-level DP clipping and Gaussian noise. & pp.~4--6, Figs.~3, 10--11.\\
SADBA~\cite{feng2025sadba} & Post-injection trajectories through 50 rounds. & Defense-aware objective references Activation Clustering~\cite{chen2018activation}/Neural Cleanse~\cite{wang2019neuralcleanse}; no comprehensive defense benchmark reported. & pp.~3--6, Figs.~5--6; client-selection sentence conflicts with large setting.\\
DMBA~\cite{liu2024dmba} & ASR-30 persistence metric; attack intervals 2/5/10/random tested. & ClippedClustering~\cite{li2023clippedclustering}, DP-FedAvg~\cite{mcmahan2018dpfedavg}, SignGuard~\cite{xu2022signguard}. & \S\S{}V-A--F, Fig. 7.\\
DBA-DFL~\cite{liu2026dbadfl} & Post-removal persistence NR. & FLIP~\cite{zhang2023flip} and FedGame~\cite{jia2023fedgame} under SWIFT~\cite{bornstein2023swift}/DSGD~\cite{lian2017dpsgd}. & \S\S{}3.1--3.2, 5.\\
FTDBA~\cite{wang2025ftdba} & Sparse attacker participation tested; persistence after removal not established. & FLTrust, FLAME, activation clustering; other tables do not isolate a named defense. & \S{}VI-B, Table IV and Table XII.\\
FDBA~\cite{wang2026fdba} & Attack-stop and removal persistence NR. & Multi-Krum~\cite{blanchard2017krum}, Trimmed Mean~\cite{yin2018trimmed}, differential privacy, FedRecover~\cite{cao2023fedrecover}. & \S\S{}6.3--6.5, Tables 5,8.\\
CollaPois~\cite{collapois2025} & Client-level ASR averaged equally over clients; no extracted post-removal horizon. & DP, norm bound~\cite{sun2019normbound}, Krum, RLR~\cite{ozdayi2021rlr}; FedAvg and personalized-FL baselines. & \S\S{}III--V, App. A/G; client-level denominator retained.\\
TFI~\cite{wang2026structure} & Temporal coordination studied; post-removal persistence NR. & Krum and DP-noise comparisons; TFI reports 58.7\% ASR at DP $\sigma=0.10$ vs 89.2\% without defense. & \S\S{}6.3, 6.5; Table 2.\\
\bottomrule
\end{tabularx}
\end{table*}

\begin{table*}[!htbp]
\caption{Conditions of graph, temporal, and adaptive-trigger studies, and one inference-time boundary case (NI-GDBA).}\label{tab:expanded-studies}
\centering\footnotesize\setlength{\tabcolsep}{4pt}
\begin{tabularx}{\textwidth}{P{.14\textwidth}P{.24\textwidth}P{.24\textwidth}Y}
\toprule
Study & Construction and task & Participation and budget & Schedule, defense tests, and source anchors\\
\midrule
Graph DBA~\cite{Xu2024FedGNNBackdoors} & Local subgraph triggers compose a global trigger; graph classification & Non-IID scenarios: $K=5$, $M=2$ or 3; additional TRIANGLES client-scale experiments & Multiple-shot; FLAME, FoolsGold, RLR. Journal \S\S3.2--3.3, 5--6, Table~2.\\
NI-GDBA~\cite{li2025nigdba} (boundary) & Clean federated training, then generators learned from the trained GNN; composite test-time graph perturbation & Default $C=5$, $c_m=2$; node/edge fractions .5/.3 and local-data fraction .5. Limited-participant settings differ & 40 FL iterations; FoolsGold tested. No training-stage implantation step in the algorithm. \S4.1, Tables 1, 3, 5--6.\\
Time Bandits~\cite{abad2024timebandits} & Temporal trigger slices across malicious devices; N-MNIST, N-CIFAR10 and Gesture & 10/25/50 devices; participation .1/.5/1; 2-attacker and 4/8-attacker settings; poison ratio $\epsilon=.1$ in Tables 3--5 & Early-only attacker selection can lose the backdoor; STRIP~\cite{gao2019strip} evaluated. \S\S3--8, Tables 2--9. This concerns input time bins as well as device allocation.\\
Improved DBA~\cite{wang2025improved} & Density-adaptive sub-triggers; projected update control; COCO, VOC, ImageNet & Protocol unresolved: the text states 120 clients and 6 selected, but the 5\%-of-selected malicious cap conflicts with the required fragment count. Actual malicious count, $r$ and $I$ remain unspecified & Krum, RFA, FoolsGold, RLR, FLAME, CRFL; FedAvg, FedProx~\cite{li2020fedprox}, and FedNova~\cite{wang2020fednova} settings. \S\S{}III--V; Tables 4--6.\\
Opt-GDBA~\cite{yang2024optgdba} & Optimized graph-trigger locations and shapes; GIN & 40 clients (20 MUTAG); 50\% sampled/round; 20\% malicious; 50\% poisoned training graphs & 200 rounds (400 RDT-M5K); empirical and certified-defense study. \S4.1.\\
\bottomrule
\end{tabularx}
\end{table*}

\begin{table*}[!htbp]
\caption{Mechanisms, participation, and outcomes of six coordinated attack studies. CoBA is coded from its journal version. Client counts are stated explicitly because source symbols differ.}\label{tab:new-fulltexts}
\centering\scriptsize\setlength{\tabcolsep}{3pt}
\begin{tabularx}{\textwidth}{P{.14\textwidth}P{.25\textwidth}P{.28\textwidth}Y}
\toprule
Study & Mechanism and task & Participation, poison budget, and schedule & Outcome, defense test, and source anchor\\
\midrule
Cooperative DBA (R-DBA)~\cite{yang2022cooperative} & Split triggers; separate clean/poison local training; jointly rotate and scale updates. MNIST, Fashion-MNIST; non-IID $\alpha=.5$. & 100 total, 10 selected, all 4 attackers forced into each round; poison fraction .3125, 10 local poison epochs. 60 global rounds; honest pretraining 10/40 rounds by task. & FLAME throughout; reduced detection with high backdoor convergence in the reported curves. No dedicated post-stop horizon. \S\S{}III--IV, Tables I--II, Figs.~5--7.\\
CoBA~\cite{lyu2025coba} & Common trigger optimized with local models; drift/diversity penalties and PGD. CIFAR-10/100, MNIST, Fashion-MNIST, LOAN. & 100 image clients/80 LOAN; 10 randomly selected per round; default 20\% controlled. Persistent and ASR-threshold intermittent modes. Default poison ratio and architecture deferred to online appendices. & 15 defenses; reported post-stop ASR above 60\% through 100 normal iterations in the persistence test. Intermittent attack frequency is a separate measure. \S\S{}III--V, Tables I--V.\\
IoT CBA~\cite{alharbi2024cba} & Local poisoning plus coordinated update perturbations; Gram--Schmidt estimation. STL-10, CIFAR-10, T-LESS, FedEMNIST; 5-layer CNN. & 40/40/40/80 clients; main table full participation and 20\% malicious. Poison 128/256 in first three tasks, 32/64 in FedEMNIST. Clean rounds 1--500, attacks 501--600; separate participation sweep 1 to .3. & Norm clipping with noise, geometric median, RLR; split-trigger CIFAR-10 variant. T-LESS industrial images; runtime report lacks hardware details. Post-stop horizon NR. \S\S{}IV--VI, Tables IV--V.\\
SSRDBA~\cite{yang2025ssrdba} & Overlapping cyclic-window shares, threshold $t=n-m+1$ for trigger coverage; passive/active rebels. FMNIST, CIFAR-10, FEDEMNIST. & Total 30/40/3383; selected 30/40/33. DBA/SSRDBA malicious counts 12/7, 16/9, 12/7 under trigger-size matching. Poison fraction unresolved: Table 2=.5 versus \S4.1=10\%. & Rebel stress tests; RFA, FoolsGold, clipping, CRFL; a mitigation proposed by the authors. No benign-only post-stop horizon. \S\S{}3--7, Theorem~1, Tables 2--4, 6.\\
SATA~\cite{ai2025sata} & Trigger splitting, probabilistic fragment association, norm control and progressive injection; CIFAR-10/ResNet-18; MNIST extension. & 50 clients, 5 malicious, 10 randomly selected per round; 4 trigger fragments, $p_{\rm assoc}=.5$; initial 20-round ramp. Local-sample poison fraction NR. Comparison: 200 attack rounds + 100 benign rounds. & ASR-100=86.68\% in its comparison (best of 3 runs); Multi-Krum and FoolsGold tested. Separate Fig.5 attack duration differs between text/caption. \S\S{}III--V, Tables II--III.\\
Dynamic\-Triggers~\cite{wang2024dynamic} & LoG image-conditioned trigger, four-part split and MixUp; ImageNet, COCO, VOC; ResNet listed. & Absolute client/selected/malicious counts NR; malicious:benign ratio 1:100. Table 1 reports 1--5\% toxicity and +.01/round without a complete budget specification. & FoolsGold; multi-shot and single-shot curves, with a 50-round single-shot continuation. Method uses LoG, while conclusion mentions an unspecified auto-encoder. \S\S{}3--4, Table~1, Figs.~4--8.\\
\bottomrule
\end{tabularx}
\end{table*}

\subsection{Source versions and reports not assessed}\label{app:evidence}
Table~\ref{tab:source-register} identifies the version of each study used for extraction. The coding of the downstream audit is in the artifact (Appendix~\ref{app:availability}).
\begin{table*}[!htbp]
\caption{Technical sources and extraction anchors for the main comparison.}\label{tab:source-register}
\centering\footnotesize\setlength{\tabcolsep}{4pt}
\begin{tabularx}{\textwidth}{P{.16\textwidth}P{.25\textwidth}Y}
\toprule
Study & Version/source used & Extraction anchor and version issue\\
\midrule
DBA & OpenReview submission and ICLR 2020 camera-ready & The two versions differ in the Table~1 poison-ratio column and one robust-aggregation table entry; camera-ready values are used.\\
FCBA & AAAI 2024 proceedings & Experiment Setup and Table 3; poisoning budgets and ASR after 120 benign rounds. Four primitives yield 14 nonempty proper subsets.\\
SADBA & AAAI 2025 proceedings & Experimental settings and Table 3; persistence plots. The large-client setup is not fully resolved: the paper gives 100/500 clients but the sampling paragraph still specifies 10 selected from 100.\\
DMBA & arXiv:2411.03926v3 & Experimental setup, attack-interval and replay ablations; ASR-30.\\
DBA-DFL & Publisher full text, ECAI 2025 lineage & Experimental settings, ring/grid/clique-ring comparisons, DSGD/SWIFT.\\
FTDBA & arXiv:2511.09252v2 & Sections III--VIII; Tables I, IV, XII. The body uses the name FracBA.\\
FDBA & arXiv:2609.07147v1 & Sections 3--7; Tables 3, 5, 8; Appendix A assumptions.\\
CollaPois & arXiv:2504.12875v2 & Experimental setup and client-level assessment; 1\% participation, attacker-share/data-heterogeneity sweeps. The ICDCS 2025 version corresponds to v2.\\
TFI & arXiv:2603.03865v1 & Experimental setup and architecture/defense comparisons; 100 clients and 10\% sampling.\\
Graph DBA & Final DTRP 2024 article & Sections 3.2--3.3, 5--6; Table 2. Explicitly extends ACSAC 2022; one study lineage.\\
CerP / CoBA & AAAI 2023 conference version and IEEE TDSC journal version & The journal version explicitly extends CerP; we use journal \S\S{}III--V/Tables I--V for updated conditions, not predecessor defaults.\\
Time Bandits & arXiv:2402.02886v1 & Sections 3--8, Tables 2--9. A later version appeared at ESORICS 2025.\\
Improved DBA & IEEE Access article & Sections III--V; Tables 4--6. The six-selected-client sentence conflicts with the stated 5\%-of-selected attacker cap and required fragments.\\
Opt-GDBA & Author manuscript of CCS 2024 paper & Section 4.1.\\
Cooperative DBA & IEEE MSN 2022 paper & \S\S{}III--IV; Tables I--II, Figs.~5--7.\\
IoT CBA & IEEE IoTJ 2024 article & \S\S{}IV--VI, Tables IV--V.\\
SSRDBA & ACM TOPS 2025 article & \S3.2/Theorem 1; \S4.1, Tables 2--4; \S\S{}5--7/Table 6. Poison-ratio table/text discrepancy retained.\\
SATA & IEEE TrustCom 2025 paper & \S\S{}III--V, Tables II--III, Figs.5--6,9; best-of-three comparison and Fig.5 duration discrepancy retained.\\
Dynamic\-Triggers & Springer WISTP 2024 chapter & \S\S{}3--4, Table 1, Figs.4--8; LoG technical mechanism; conclusion's auto-encoder description unresolved.\\
\bottomrule
\end{tabularx}
\end{table*}

Eight attack-relevant reports could not be assessed in full text before the cutoff. Seven lacked an accessible complete technical version, and one (MABA, issue date 15 December 2026) also lacks a confirmed pre-cutoff technical version. We did not use the publication year alone as the cutoff test. For example, PFed-GDBA carries a 2027 volume year but was first published online on 13 July 2026, so it is within scope pending full-text access. These reports will be coded when their full texts become available.

\section{Additional Results and Protocol Details}\label{app:experiment}
Tables~\ref{tab:emp-paired}, \ref{tab:emp-controls}, and~\ref{tab:emp-controls2} give the values behind Figs.~\ref{fig:emp-effects} and~\ref{fig:emp-conditions}, Table~\ref{tab:emp-nt} the Neurotoxin comparison, and Table~\ref{tab:emp-repair} the server-side repairs.

\begin{table*}[!htbp]
\caption{Paired differences in ASR of the full trigger under random participation (percentage points), at the end of the attack window ($t=0$) and after 100 benign rounds ($t=100$). Each cell gives the mean over five paired seeds and, below it, the range of the per-seed differences. Bold: all five seeds share the sign. Resource: B$-$A; composition: C$-$B (B and C are matched in submitted updates, poisoned samples, and aggregation weight); redundancy: D$-$C.}\label{tab:emp-paired}
\centering\footnotesize\setlength{\tabcolsep}{3pt}
\begin{tabular}{llcccccc}
\toprule
& & \multicolumn{2}{c}{Resource (B$-$A)} & \multicolumn{2}{c}{Composition (C$-$B)} & \multicolumn{2}{c}{Redundancy (D$-$C)}\\
\cmidrule(lr){3-4}\cmidrule(lr){5-6}\cmidrule(lr){7-8}
Dataset & Model & $t=0$ & $t=100$ & $t=0$ & $t=100$ & $t=0$ & $t=100$\\
\midrule
MNIST & MLP & \makecell{\textbf{12.6}\\{\tiny 1.4 to 33.0}} & \makecell{\textbf{15.9}\\{\tiny 2.5 to 35.2}} & \makecell{\textbf{0.7}\\{\tiny 0.1 to 1.6}} & \makecell{\textbf{2.5}\\{\tiny 0.5 to 4.3}} & \makecell{0.0\\{\tiny 0.0 to 0.0}} & \makecell{0.0\\{\tiny 0.0 to 0.0}}\\[2pt]
MNIST & CNN & \makecell{\textbf{50.5}\\{\tiny 19.7 to 79.1}} & \makecell{\textbf{48.9}\\{\tiny 33.7 to 61.3}} & \makecell{-56.1\\{\tiny -95.1 to 2.2}} & \makecell{-38.3\\{\tiny -70.2 to 16.8}} & \makecell{42.8\\{\tiny -0.6 to 96.9}} & \makecell{35.1\\{\tiny -10.1 to 87.3}}\\[2pt]
FMNIST & MLP & \makecell{\textbf{33.9}\\{\tiny 10.9 to 73.3}} & \makecell{\textbf{33.8}\\{\tiny 11.3 to 66.9}} & \makecell{\textbf{1.6}\\{\tiny 0.4 to 3.0}} & \makecell{\textbf{2.1}\\{\tiny 0.9 to 3.4}} & \makecell{0.5\\{\tiny -0.0 to 2.2}} & \makecell{1.5\\{\tiny -0.0 to 3.7}}\\[2pt]
FMNIST & CNN & \makecell{\textbf{64.6}\\{\tiny 44.6 to 91.4}} & \makecell{\textbf{60.3}\\{\tiny 52.3 to 68.9}} & \makecell{\textbf{-53.8}\\{\tiny -87.6 to -11.2}} & \makecell{\textbf{-54.0}\\{\tiny -83.1 to -22.1}} & \makecell{\textbf{53.7}\\{\tiny 12.3 to 88.9}} & \makecell{\textbf{54.3}\\{\tiny 21.2 to 84.2}}\\[2pt]
CIFAR-10 & ResNet-18 & \makecell{\textbf{35.5}\\{\tiny 33.1 to 39.8}} & \makecell{\textbf{36.0}\\{\tiny 20.9 to 48.2}} & \makecell{-0.6\\{\tiny -5.4 to 4.3}} & \makecell{-9.6\\{\tiny -26.2 to 4.3}} & \makecell{\textbf{2.1}\\{\tiny 0.3 to 5.8}} & \makecell{\textbf{13.0}\\{\tiny 6.4 to 23.9}}\\[2pt]
CIFAR-10 & VGG-11 & \makecell{\textbf{34.8}\\{\tiny 22.4 to 43.9}} & \makecell{\textbf{37.1}\\{\tiny 27.2 to 45.2}} & \makecell{\textbf{-5.7}\\{\tiny -10.5 to -0.6}} & \makecell{\textbf{-16.5}\\{\tiny -22.9 to -10.6}} & \makecell{\textbf{7.0}\\{\tiny 2.1 to 11.2}} & \makecell{\textbf{14.4}\\{\tiny 1.3 to 24.5}}\\[2pt]
\bottomrule
\end{tabular}
\end{table*}

\begin{table}[!htbp]
\caption{Supplementary controls (paired differences in ASR of the full trigger, percentage points; mean over five seeds and, below it, the per-seed range; bold: all seeds share the sign). Pixel-matched: B-q4 and B-q2 apply the full trigger to one quarter and one half as many poisoned samples as B, matching the trigger pixels of C and D to within 4\%. DBA-style setting: forced participation, attackers inject one after another with scaled updates, $t=0$ after the last injection. Corners: the same pieces placed in the four image corners, with B, C, and B-q4 defined as before. No defense. $^{\dagger}$Four seeds; the seed-2 run of A produced non-finite outputs.}\label{tab:emp-controls}
\centering\scriptsize\setlength{\tabcolsep}{3pt}
\begin{tabular}{llcc}
\toprule
Configuration & Contrast & $t=0$ & $t=100$\\
\midrule
\multicolumn{4}{l}{\emph{Pixel-matched control (random participation)}}\\
\makecell[l]{CIFAR-10\\ResNet-18} & C$-$B (upd.) & \makecell{-0.6\\{\tiny -5.4 to 4.3}} & \makecell{-9.6\\{\tiny -26.2 to 4.3}}\\[1pt]
 & C$-$(B-q4) & \makecell{\textbf{5.8}\\{\tiny 2.3 to 9.7}} & \makecell{-1.5\\{\tiny -15.0 to 5.4}}\\[1pt]
 & D$-$(B-q2) & \makecell{\textbf{3.4}\\{\tiny 1.6 to 8.1}} & \makecell{\textbf{6.8}\\{\tiny 0.8 to 10.9}}\\[1pt]
 & B$-$(B-q4) & \makecell{\textbf{6.4}\\{\tiny 5.4 to 7.6}} & \makecell{8.2\\{\tiny -1.8 to 12.9}}\\[1pt]
\makecell[l]{FMNIST\\CNN} & C$-$B (upd.) & \makecell{\textbf{-53.8}\\{\tiny -87.6 to -11.2}} & \makecell{\textbf{-54.0}\\{\tiny -83.1 to -22.1}}\\[1pt]
 & C$-$(B-q4) & \makecell{-30.2\\{\tiny -71.3 to 3.7}} & \makecell{-22.0\\{\tiny -57.2 to 10.4}}\\[1pt]
 & D$-$(B-q2) & \makecell{\textbf{7.0}\\{\tiny 4.3 to 11.3}} & \makecell{\textbf{12.8}\\{\tiny 8.7 to 23.0}}\\[1pt]
 & B$-$(B-q4) & \makecell{\textbf{23.6}\\{\tiny 12.0 to 39.4}} & \makecell{\textbf{32.0}\\{\tiny 22.2 to 38.4}}\\[1pt]
\midrule
\multicolumn{4}{l}{\emph{DBA-style setting ($\gamma\eta/n=1$, staged single-shot)}}\\
\makecell[l]{CIFAR-10\\ResNet-18} & C$-$B & \makecell{1.8\\{\tiny 0.0 to 3.8}} & \makecell{\textbf{-12.8}\\{\tiny -21.9 to -0.3}}\\[1pt]
 & B$-$A4 & \makecell{\textbf{-3.5}\\{\tiny -4.4 to -1.6}} & \makecell{6.1\\{\tiny -2.8 to 12.1}}\\[1pt]
 & B$-$A$^{\dagger}$ & \makecell{4.0\\{\tiny -1.6 to 9.6}} & \makecell{\textbf{22.8}\\{\tiny 17.0 to 32.9}}\\[1pt]
 & C$-$A$^{\dagger}$ & \makecell{6.3\\{\tiny -0.9 to 11.6}} & \makecell{\textbf{8.5}\\{\tiny 2.7 to 16.7}}\\[1pt]
\midrule
\multicolumn{4}{l}{\emph{Pieces in the four corners (random participation)}}\\
\makecell[l]{CIFAR-10\\ResNet-18} & C$-$B & \makecell{\textbf{7.1}\\{\tiny 4.1 to 10.0}} & \makecell{\textbf{16.6}\\{\tiny 11.1 to 20.4}}\\[1pt]
 & C$-$(B-q4) & \makecell{\textbf{17.8}\\{\tiny 13.3 to 23.4}} & \makecell{\textbf{29.0}\\{\tiny 25.9 to 33.3}}\\[1pt]
\makecell[l]{FMNIST\\CNN} & C$-$B & \makecell{\textbf{-62.9}\\{\tiny -79.5 to -17.8}} & \makecell{\textbf{-48.4}\\{\tiny -59.9 to -34.6}}\\[1pt]
 & C$-$(B-q4) & \makecell{-10.7\\{\tiny -28.1 to 6.3}} & \makecell{-5.5\\{\tiny -17.6 to 5.6}}\\[1pt]
\makecell[l]{CIFAR-10\\VGG-11} & C$-$B & \makecell{\textbf{5.8}\\{\tiny 2.6 to 9.0}} & \makecell{\textbf{10.0}\\{\tiny 6.9 to 14.5}}\\[1pt]
\makecell[l]{MNIST\\CNN} & C$-$B & \makecell{\textbf{-63.0}\\{\tiny -92.8 to -0.1}} & \makecell{\textbf{-35.1}\\{\tiny -70.0 to -0.0}}\\[1pt]
\midrule
\multicolumn{4}{l}{\emph{Bars 8 pixels apart (random participation)}}\\
\makecell[l]{CIFAR-10\\ResNet-18} & C$-$B & \makecell{1.9\\{\tiny -0.4 to 4.5}} & \makecell{3.8\\{\tiny -1.4 to 16.3}}\\[1pt]
\bottomrule
\end{tabular}
\end{table}

\begin{table}[!htbp]
\caption{Second-round supplementary controls, as in Table~\ref{tab:emp-controls}. Further datasets use ResNet-18 with their calibrated $E_{\mathrm{adv}}$ (6, 4, and 2 for grayscale CIFAR-10, GTSRB, and FMNIST). FCBA14 gives each of 14 attackers a different nonempty proper subset of the pieces; B14 and B14-q2 use $G$ with the same and half the poisoned samples. No defense.}\label{tab:emp-controls2}
\centering\scriptsize\setlength{\tabcolsep}{3pt}
\begin{tabular}{llcc}
\toprule
Configuration & Contrast & $t=0$ & $t=100$\\
\midrule
\multicolumn{4}{l}{\emph{Further datasets, ResNet-18 (random participation)}}\\
\makecell[l]{Gray\\CIFAR-10} & C$-$B & \makecell{-1.2\\{\tiny -3.6 to 0.6}} & \makecell{-7.0\\{\tiny -18.6 to 8.9}}\\[1pt]
 & C$-$B (corners) & \makecell{\textbf{6.8}\\{\tiny 3.5 to 13.1}} & \makecell{\textbf{18.5}\\{\tiny 12.9 to 24.1}}\\[1pt]
\makecell[l]{GTSRB} & B$-$A & \makecell{\textbf{8.3}\\{\tiny 4.3 to 10.0}} & \makecell{\textbf{12.5}\\{\tiny 6.2 to 20.2}}\\[1pt]
 & C$-$B & \makecell{\textbf{-1.2}\\{\tiny -1.8 to -0.5}} & \makecell{\textbf{-1.3}\\{\tiny -1.8 to -0.6}}\\[1pt]
 & C$-$B (corners) & \makecell{\textbf{1.5}\\{\tiny 0.9 to 2.2}} & \makecell{\textbf{2.6}\\{\tiny 1.2 to 3.5}}\\[1pt]
\makecell[l]{FMNIST} & C$-$B & \makecell{\textbf{0.1}\\{\tiny 0.0 to 0.1}} & \makecell{0.1\\{\tiny 0.0 to 0.2}}\\[1pt]
 & C$-$B (corners) & \makecell{\textbf{0.2}\\{\tiny 0.0 to 0.3}} & \makecell{\textbf{0.4}\\{\tiny 0.1 to 1.0}}\\[1pt]
\midrule
\multicolumn{4}{l}{\emph{FCBA-style allocation, 14 attackers (random participation)}}\\
\makecell[l]{CIFAR-10\\ResNet-18} & B14$-$A & \makecell{\textbf{48.7}\\{\tiny 29.9 to 65.7}} & \makecell{\textbf{58.2}\\{\tiny 45.4 to 62.8}}\\[1pt]
 & FCBA14$-$B14 & \makecell{\textbf{1.0}\\{\tiny 0.1 to 1.7}} & \makecell{\textbf{9.1}\\{\tiny 2.9 to 16.3}}\\[1pt]
 & FCBA14$-$(B14-q2) & \makecell{\textbf{2.5}\\{\tiny 1.1 to 3.8}} & \makecell{\textbf{8.8}\\{\tiny 2.9 to 15.9}}\\[1pt]
\makecell[l]{FMNIST\\CNN} & B14$-$A & \makecell{\textbf{74.5}\\{\tiny 47.9 to 98.2}} & \makecell{\textbf{83.4}\\{\tiny 68.1 to 96.4}}\\[1pt]
 & FCBA14$-$B14 & \makecell{\textbf{0.2}\\{\tiny 0.1 to 0.4}} & \makecell{\textbf{1.9}\\{\tiny 0.9 to 2.9}}\\[1pt]
 & FCBA14$-$(B14-q2) & \makecell{\textbf{0.3}\\{\tiny 0.1 to 0.8}} & \makecell{\textbf{3.1}\\{\tiny 1.4 to 4.5}}\\[1pt]
\midrule
\multicolumn{4}{l}{\emph{Larger $E_{\mathrm{adv}}$, compact layout (random participation)}}\\
\makecell[l]{MNIST\\CNN} & C$-$B, $E_{\mathrm{adv}}=4$ & \makecell{-11.2\\{\tiny -69.7 to 6.8}} & \makecell{-2.4\\{\tiny -67.3 to 20.3}}\\[1pt]
 & C$-$B, $E_{\mathrm{adv}}=6$ & \makecell{1.6\\{\tiny -3.9 to 5.8}} & \makecell{8.7\\{\tiny -21.9 to 21.3}}\\[1pt]
\makecell[l]{FMNIST\\CNN} & C$-$B, $E_{\mathrm{adv}}=4$ & \makecell{0.3\\{\tiny -1.2 to 1.7}} & \makecell{0.7\\{\tiny -4.9 to 4.3}}\\[1pt]
 & C$-$B, $E_{\mathrm{adv}}=6$ & \makecell{0.8\\{\tiny -0.2 to 2.8}} & \makecell{\textbf{4.9}\\{\tiny 0.5 to 16.7}}\\[1pt]
\bottomrule
\end{tabular}
\end{table}

Table~\ref{tab:emp-asr} lists the ASR of the full trigger for every arm, and Table~\ref{tab:emp-paired-forced} gives the paired differences under forced participation. Per-seed values for all 15 piece subsets, the update statistics, and the resource audit are in the artifact.

\begin{table*}[!htbp]
\caption{ASR of the full trigger (percent, mean $\pm$ standard deviation over five seeds) at the end of the attack window ($t=0$) and after 100 benign rounds ($t=100$), without defense.}\label{tab:emp-asr}
\centering\scriptsize\setlength{\tabcolsep}{2pt}
\resizebox{\textwidth}{!}{%
\begin{tabular}{llcccccccc}
\toprule
& & \multicolumn{4}{c}{$t=0$} & \multicolumn{4}{c}{$t=100$}\\
\cmidrule(lr){3-6}\cmidrule(lr){7-10}
Dataset & Model & A & B & C & D & A & B & C & D\\
\midrule
\multicolumn{10}{l}{\emph{Random participation (200-round window)}}\\
MNIST & MLP & 86.7$\pm$12.6 & 99.3$\pm$0.6 & 100.0$\pm$0.0 & 100.0$\pm$0.0 & 81.6$\pm$13.0 & 97.5$\pm$1.4 & 100.0$\pm$0.0 & 100.0$\pm$0.0\\
MNIST & CNN & 41.5$\pm$24.6 & 92.0$\pm$4.7 & 36.0$\pm$49.2 & 78.8$\pm$40.5 & 18.4$\pm$16.9 & 67.3$\pm$11.1 & 29.0$\pm$43.4 & 64.2$\pm$37.5\\
FMNIST & MLP & 63.9$\pm$29.1 & 97.8$\pm$2.2 & 99.4$\pm$1.1 & 99.9$\pm$0.2 & 62.3$\pm$27.8 & 96.1$\pm$3.2 & 98.2$\pm$2.4 & 99.7$\pm$0.6\\
FMNIST & CNN & 29.9$\pm$26.2 & 94.4$\pm$4.9 & 40.6$\pm$30.8 & 94.3$\pm$6.1 & 19.0$\pm$16.6 & 79.3$\pm$13.0 & 25.3$\pm$22.3 & 79.6$\pm$16.3\\
CIFAR-10 & ResNet-18 & 51.5$\pm$7.3 & 87.0$\pm$4.7 & 86.4$\pm$4.5 & 88.4$\pm$2.6 & 17.2$\pm$5.4 & 53.2$\pm$11.5 & 43.6$\pm$4.1 & 56.6$\pm$5.1\\
CIFAR-10 & VGG-11 & 46.3$\pm$11.1 & 81.1$\pm$3.5 & 75.4$\pm$7.4 & 82.4$\pm$4.0 & 23.7$\pm$8.7 & 60.8$\pm$3.3 & 44.3$\pm$7.4 & 58.8$\pm$3.5\\
\midrule
\multicolumn{10}{l}{\emph{Forced participation (20-round window)}}\\
MNIST & MLP & 92.1$\pm$7.2 & 99.6$\pm$0.5 & 100.0$\pm$0.0 & 100.0$\pm$0.0 & 80.3$\pm$11.8 & 95.1$\pm$3.0 & 99.9$\pm$0.2 & 99.9$\pm$0.2\\
MNIST & CNN & 63.1$\pm$15.5 & 95.8$\pm$4.6 & 42.2$\pm$52.7 & 93.0$\pm$14.2 & 12.9$\pm$14.0 & 30.0$\pm$16.9 & 23.5$\pm$34.8 & 44.7$\pm$28.7\\
FMNIST & MLP & 76.7$\pm$12.3 & 99.8$\pm$0.1 & 100.0$\pm$0.0 & 100.0$\pm$0.0 & 53.0$\pm$19.2 & 85.1$\pm$8.1 & 97.6$\pm$2.7 & 98.0$\pm$1.5\\
FMNIST & CNN & 41.8$\pm$33.9 & 98.5$\pm$0.3 & 78.0$\pm$38.4 & 99.5$\pm$0.1 & 11.5$\pm$9.5 & 43.7$\pm$14.0 & 22.5$\pm$13.9 & 55.7$\pm$13.7\\
CIFAR-10 & ResNet-18 & 74.0$\pm$6.6 & 92.0$\pm$0.9 & 92.4$\pm$1.2 & 93.0$\pm$1.0 & 15.8$\pm$10.8 & 32.0$\pm$12.7 & 26.0$\pm$7.6 & 36.1$\pm$9.6\\
CIFAR-10 & VGG-11 & 67.2$\pm$3.6 & 87.6$\pm$2.1 & 87.2$\pm$1.6 & 87.9$\pm$2.2 & 15.8$\pm$10.0 & 35.1$\pm$12.8 & 27.3$\pm$12.9 & 38.8$\pm$9.4\\
\bottomrule
\end{tabular}%
}
\end{table*}

\begin{table*}[!htbp]
\caption{Paired differences under forced participation (percentage points; mean over five paired seeds and range of the per-seed differences). Bold: all five seeds share the sign.}\label{tab:emp-paired-forced}
\centering\footnotesize\setlength{\tabcolsep}{3pt}
\begin{tabular}{llcccccc}
\toprule
& & \multicolumn{2}{c}{Resource (B$-$A)} & \multicolumn{2}{c}{Composition (C$-$B)} & \multicolumn{2}{c}{Redundancy (D$-$C)}\\
\cmidrule(lr){3-4}\cmidrule(lr){5-6}\cmidrule(lr){7-8}
Dataset & Model & $t=0$ & $t=100$ & $t=0$ & $t=100$ & $t=0$ & $t=100$\\
\midrule
MNIST & MLP & \makecell{\textbf{7.5}\\{\tiny 0.7 to 18.3}} & \makecell{\textbf{14.8}\\{\tiny 4.3 to 32.0}} & \makecell{0.4\\{\tiny 0.0 to 1.3}} & \makecell{\textbf{4.8}\\{\tiny 0.9 to 9.0}} & \makecell{0.0\\{\tiny 0.0 to 0.0}} & \makecell{-0.0\\{\tiny -0.2 to 0.0}}\\[2pt]
MNIST & CNN & \makecell{\textbf{32.6}\\{\tiny 11.7 to 52.8}} & \makecell{\textbf{17.1}\\{\tiny 5.6 to 25.7}} & \makecell{-53.6\\{\tiny -94.8 to 1.5}} & \makecell{-6.5\\{\tiny -31.2 to 22.7}} & \makecell{50.9\\{\tiny -0.1 to 96.1}} & \makecell{\textbf{21.1}\\{\tiny 1.5 to 50.2}}\\[2pt]
FMNIST & MLP & \makecell{\textbf{23.1}\\{\tiny 9.3 to 36.6}} & \makecell{\textbf{32.2}\\{\tiny 16.3 to 48.4}} & \makecell{\textbf{0.2}\\{\tiny 0.1 to 0.3}} & \makecell{\textbf{12.5}\\{\tiny 5.3 to 19.9}} & \makecell{-0.0\\{\tiny -0.0 to 0.0}} & \makecell{0.4\\{\tiny -1.1 to 2.4}}\\[2pt]
FMNIST & CNN & \makecell{\textbf{56.7}\\{\tiny 18.9 to 95.9}} & \makecell{\textbf{32.1}\\{\tiny 19.4 to 43.1}} & \makecell{\textbf{-20.5}\\{\tiny -88.1 to -0.3}} & \makecell{-21.2\\{\tiny -55.2 to 1.7}} & \makecell{\textbf{21.5}\\{\tiny 1.1 to 89.7}} & \makecell{\textbf{33.2}\\{\tiny 17.4 to 64.3}}\\[2pt]
CIFAR-10 & ResNet-18 & \makecell{\textbf{18.0}\\{\tiny 8.6 to 24.7}} & \makecell{\textbf{16.2}\\{\tiny 8.2 to 29.2}} & \makecell{0.5\\{\tiny -1.4 to 1.6}} & \makecell{-6.0\\{\tiny -12.9 to 6.9}} & \makecell{\textbf{0.5}\\{\tiny 0.1 to 0.8}} & \makecell{\textbf{10.1}\\{\tiny 4.4 to 14.6}}\\[2pt]
CIFAR-10 & VGG-11 & \makecell{\textbf{20.3}\\{\tiny 17.2 to 23.9}} & \makecell{\textbf{19.3}\\{\tiny 11.4 to 25.0}} & \makecell{-0.4\\{\tiny -2.9 to 1.0}} & \makecell{-7.7\\{\tiny -13.0 to 5.6}} & \makecell{0.7\\{\tiny -1.1 to 2.7}} & \makecell{\textbf{11.5}\\{\tiny 6.1 to 15.8}}\\[2pt]
\bottomrule
\end{tabular}
\end{table*}

\begin{table*}[!htbp]
\caption{Neurotoxin single attacker (A-NT) under random participation without defense: paired differences in ASR of the full trigger (percentage points; mean over five seeds and per-seed range; bold: all five seeds share the sign). A-NT masks the 5\% largest coordinates of the previous global update during local training; all other settings equal those of arm A.}\label{tab:emp-nt}
\centering\footnotesize\setlength{\tabcolsep}{3pt}
\begin{tabular}{llcccccc}
\toprule
& & \multicolumn{2}{c}{A-NT$-$A} & \multicolumn{2}{c}{B$-$A-NT} & \multicolumn{2}{c}{C$-$A-NT}\\
\cmidrule(lr){3-4}\cmidrule(lr){5-6}\cmidrule(lr){7-8}
Dataset & Model & $t=0$ & $t=100$ & $t=0$ & $t=100$ & $t=0$ & $t=100$\\
\midrule
MNIST & MLP & \makecell{-0.1\\{\tiny -1.3 to 0.5}} & \makecell{-0.1\\{\tiny -2.2 to 0.9}} & \makecell{\textbf{12.7}\\{\tiny 1.7 to 33.0}} & \makecell{\textbf{16.1}\\{\tiny 2.8 to 35.0}} & \makecell{\textbf{13.4}\\{\tiny 1.9 to 33.8}} & \makecell{\textbf{18.6}\\{\tiny 3.3 to 37.9}}\\[2pt]
MNIST & CNN & \makecell{\textbf{-9.4}\\{\tiny -14.7 to -2.3}} & \makecell{\textbf{-4.2}\\{\tiny -11.0 to -0.9}} & \makecell{\textbf{59.9}\\{\tiny 34.4 to 81.4}} & \makecell{\textbf{53.1}\\{\tiny 43.0 to 63.8}} & \makecell{3.8\\{\tiny -31.5 to 72.1}} & \makecell{14.8\\{\tiny -17.0 to 61.5}}\\[2pt]
FMNIST & MLP & \makecell{\textbf{-1.1}\\{\tiny -2.8 to -0.2}} & \makecell{-0.8\\{\tiny -2.4 to 0.5}} & \makecell{\textbf{35.0}\\{\tiny 11.3 to 73.9}} & \makecell{\textbf{34.5}\\{\tiny 12.5 to 67.4}} & \makecell{\textbf{36.5}\\{\tiny 11.6 to 77.0}} & \makecell{\textbf{36.6}\\{\tiny 13.4 to 70.8}}\\[2pt]
FMNIST & CNN & \makecell{\textbf{-6.8}\\{\tiny -13.7 to -0.1}} & \makecell{\textbf{-4.1}\\{\tiny -8.7 to -0.1}} & \makecell{\textbf{71.4}\\{\tiny 54.7 to 91.7}} & \makecell{\textbf{64.3}\\{\tiny 59.1 to 69.0}} & \makecell{17.6\\{\tiny -24.6 to 53.6}} & \makecell{10.4\\{\tiny -20.3 to 40.2}}\\[2pt]
CIFAR-10 & ResNet-18 & \makecell{\textbf{-8.3}\\{\tiny -10.8 to -2.5}} & \makecell{\textbf{-3.8}\\{\tiny -5.9 to -1.6}} & \makecell{\textbf{43.7}\\{\tiny 36.1 to 50.6}} & \makecell{\textbf{39.8}\\{\tiny 25.7 to 54.0}} & \makecell{\textbf{43.1}\\{\tiny 35.3 to 51.7}} & \makecell{\textbf{30.2}\\{\tiny 27.3 to 33.2}}\\[2pt]
CIFAR-10 & VGG-11 & \makecell{\textbf{-7.3}\\{\tiny -13.0 to -3.0}} & \makecell{\textbf{-4.8}\\{\tiny -7.2 to -2.8}} & \makecell{\textbf{42.1}\\{\tiny 27.9 to 51.0}} & \makecell{\textbf{41.9}\\{\tiny 34.4 to 47.9}} & \makecell{\textbf{36.4}\\{\tiny 27.3 to 43.4}} & \makecell{\textbf{25.4}\\{\tiny 17.5 to 34.3}}\\[2pt]
\bottomrule
\end{tabular}
\end{table*}

\begin{table*}[!htbp]
\caption{Server-side repairs of retrained CIFAR-10 models at the end of the attack window (percent, mean over five seeds; random participation, no defense during training). CA and ASR($G$) before$\to$after the repair; ``piece'': best single-piece ASR after the repair. All methods use FLTrust's 100 root images and hyperparameters fixed on seed 0. No model of either arm met the missed-piece criterion (ASR($G$)${}<20\%$ and a piece $\geq 50\%$); per-seed values and the reverse-engineered masks are in the artifact.}\label{tab:emp-repair}
\centering\footnotesize\setlength{\tabcolsep}{4pt}
\begin{tabular}{llcccccc}
\toprule
& & \multicolumn{3}{c}{B (full trigger)} & \multicolumn{3}{c}{C (pieces)}\\
\cmidrule(lr){3-5}\cmidrule(lr){6-8}
Configuration & Repair & CA & ASR($G$) & Piece & CA & ASR($G$) & Piece\\
\midrule
ResNet-18, compact & Fine-pruning & 85.7$\to$80.4 & 87.0$\to$85.6 & 20.4 & 85.9$\to$80.7 & 86.4$\to$85.9 & 70.5\\
 & Fine-tuning & 85.7$\to$80.5 & 87.0$\to$86.0 & 20.8 & 85.9$\to$80.6 & 86.4$\to$86.4 & 71.7\\
 & Neural Cleanse & 85.7$\to$79.2 & 87.0$\to$20.9 & 1.3 & 85.9$\to$78.4 & 86.4$\to$11.7 & 4.2\\
\midrule
ResNet-18, corners & Fine-pruning & 86.1$\to$80.4 & 85.7$\to$85.9 & 32.1 & 85.8$\to$80.3 & 92.7$\to$92.3 & 65.0\\
 & Fine-tuning & 86.1$\to$80.3 & 85.7$\to$86.6 & 33.2 & 85.8$\to$80.4 & 92.7$\to$92.7 & 66.6\\
 & Neural Cleanse & 86.1$\to$80.0 & 85.7$\to$35.4 & 4.0 & 85.8$\to$80.3 & 92.7$\to$58.4 & 11.2\\
\midrule
VGG-11, compact & Fine-pruning & 82.4$\to$76.6 & 81.1$\to$83.3 & 34.8 & 81.9$\to$77.2 & 75.4$\to$80.2 & 72.2\\
 & Fine-tuning & 82.4$\to$76.6 & 81.1$\to$83.2 & 33.9 & 81.9$\to$77.3 & 75.4$\to$80.4 & 72.4\\
 & Neural Cleanse & 82.4$\to$72.8 & 81.1$\to$27.7 & 5.0 & 81.9$\to$73.2 & 75.4$\to$19.0 & 23.3\\
\bottomrule
\end{tabular}
\end{table*}

\subsection{Protocol details and deviations}
\paragraph{Frozen configuration} All hyperparameters were fixed before the reported seeds 1--5 were run, in the configuration file \texttt{frozen.yaml} of the artifact (SHA-256 prefix \texttt{eb974ec9d021}).

\paragraph{Calibration on seed 0} Table~\ref{tab:calibration} lists the values fixed on seed 0.

\begin{table*}[!htbp]
\caption{Calibration on seed 0. $^{\dagger}$No candidate reached the 90\% target. $^{*}$Original implementation: no candidate met the one-point clean-accuracy tolerance, and the smallest was used (Deviation 1). The corrected CIFAR-10 defense runs in Table~\ref{tab:emp-defense} use $\sigma=0.0005$ (Deviation 13).}\label{tab:calibration}
\centering\footnotesize
\begin{tabular}{llrrrrr}
\toprule
Dataset & Model & Pretraining rounds & Clean CA (\%) & $E_{\mathrm{adv}}$ & Clipping norm $C$ & Noise $\sigma$\\
\midrule
MNIST & CNN & 68 & 98.40 & 2 & 0.185 & 0.005\\
MNIST & MLP & 518 & 98.10 & 2 & 0.144 & 0.005\\
FMNIST & CNN & 90 & 86.14 & 2 & 0.425 & 0.005\\
FMNIST & MLP & 135 & 86.46 & 2 & 0.449 & 0.005\\
CIFAR-10 & ResNet-18 & 450 & 82.98 & 6 & 4.791 & 0.001$^{*}$\\
CIFAR-10 & VGG-11 & 300 & 75.80 & 6 & 6.316 & 0.001$^{*}$\\
Gray CIFAR-10 & ResNet-18 & 300 & 72.37 & 6$^{\dagger}$ & -- & --\\
FMNIST & ResNet-18 & 60 & 87.35 & 2 & -- & --\\
GTSRB & ResNet-18 & 300 & 94.72 & 4 & -- & --\\
\bottomrule
\end{tabular}
\end{table*}

$E_{\mathrm{adv}}$ was selected on arm B under random participation. On CIFAR-10, arm B reached 72.7\%, 81.5\%, and 83.8\% ASR for $E_{\mathrm{adv}}=2$, 4, and 6. No candidate reached the 90\% target, so 6 was used. The clipping norm $C$ is the median L2 norm of all benign updates in 300 clean rounds.

\paragraph{Deviations from the written protocol}
\begin{enumerate}
\item On CIFAR-10, the noise level $\sigma=0.001$ was used although it reduced clean accuracy on seed 0 by 19.9 (ResNet-18) and 35.9 (VGG-11) points, beyond the one-point tolerance. This exception was fixed before any reported seed was run.
\item The clean arm N was run once per configuration under random participation for 300 rounds and evaluated at rounds 20, 45, 70, 120, 200, 225, 250, and 300. It serves as a shared descriptive reference and is not paired with forced participation.
\item $U$, $P$, and $\Omega$ count malicious updates as submitted to the aggregator. The number admitted by a defense and the resulting aggregation weight are recorded separately.
\item Under forced participation, the four controlled clients join every round of the window and six benign clients are added. In arm A, $m_2,\dots,m_4$ train normally. All four controlled clients are removed after the window in every arm.
\item The clipping norm and the noise level were calibrated per dataset and model on the final clean accuracy after 300 rounds. Pretraining lengths were extended by a factor of 1.5 (rounded up) until the accuracy threshold was reached on seed 0.
\item Multi-Krum scores all updates once and keeps the eight with the best scores; FLAME returns a zero update when no valid cluster exists, and FLTrust returns a zero update when all trust scores are zero.
\item Composition classes give priority to the full trigger (Table~\ref{tab:emp-classes}).
\item The pixel-matched control (arms B-q4 and B-q2; 20 runs, seeds 1--5, reusing the B, C, and D runs) was added after the main results were known, on configurations chosen in view of them. Poisoned samples are allocated per batch with a floor that resets every epoch, and the last batch is truncated to its actual size, so the trigger pixel positions of B-q4 and B-q2 differ from those of C and D by $-3.9\%$ to $+1.4\%$ per run; $U$ and $\Omega$ equal those of B exactly. All 20 runs completed.
\item The DBA-setting replication (arms A, A4, B, C; 20 runs) uses its own client-sampling stream: in each round ten benign clients are drawn, and in an injection round the scheduled attacker replaces the tenth. A injects once, in round 7, so that all arms share $t=0$. The optimizer settings are those of the main experiment (learning rate 0.01 for benign and malicious clients) rather than DBA's (0.1 and 0.05 with decay). The seed-2 run of arm A produced non-finite evaluation outputs in round 7 and was not rerun.
\item The corner layout (arms B-sep, C-sep, B-sep-q4; 30 runs) reuses $E_{\mathrm{adv}}$ from the main experiment. The trigger pixel positions of B-sep-q4 differ from those of C-sep by $-3.9\%$ to $+1.5\%$ per run.
\item The Neurotoxin attacker (30 runs) masks the largest 5\% of coordinates over all trainable parameters, following the image setting of its paper. It estimates the benign direction from the previous global update, uses no additional norm projection, and keeps the momentum and weight decay of the main experiment. In every seed the attacker was first sampled after round 5, so a previous global update was always available.
\item The separated layouts (30 runs) move the same pieces to the image corners on CIFAR-10/VGG-11 and MNIST/CNN, and to bars 8 pixels apart on CIFAR-10/ResNet-18; $E_{\mathrm{adv}}$ is reused. The corner layout also changes the absolute position of the pieces and their distance to the border, so the layout sequence on ResNet-18 is descriptive, not a one-parameter test.
\item The corrected CIFAR-10 defense runs (60 runs) exclude the batch-normalization running statistics from all defense computations and average them over the accepted clients. With this correction, the original noise candidates still cost too much clean accuracy on seed 0 (2.1, 4.7, and 16.1 points at $\sigma=0.001$, 0.002, and 0.005). The candidate set was then extended to $\{0.0005, 0.0002, 0.0001\}$ on clean runs only, and $\sigma=0.0005$ (0.8 points) was selected before any attack run. FLAME keeps its own fixed noise factor. Compared with the original runs, the correction lowered ASR under Multi-Krum and raised it under FLTrust, where the share of admitted malicious updates rose from about 26\% to about 60\%.
\item A diagnostic replay on seed 0 showed negative batch-normalization running variances in VGG-11 after 14 rounds with clipping and noise. Normalization statistics were neither clamped nor excluded from aggregation.
\item The second supplementary round (175 runs: further datasets 65, repairs 30, FCBA-style allocation 40, larger $E_{\mathrm{adv}}$ 40) was written after an internal review of an earlier draft and run without further changes to the existing results. The SHA-256 hashes of all 19{,}082 earlier files were unchanged afterwards. All 175 runs completed. For the further datasets, pretraining and $E_{\mathrm{adv}}$ were calibrated on seed 0 as in the main experiment. Grayscale CIFAR-10 converts each image with the standard luminance transform and replicates it to three channels, keeping CIFAR-10's normalization and augmentation; FMNIST is replicated to three channels without augmentation; GTSRB uses the official training set (39{,}209 images), resized to $32\times32$ without augmentation, and 430 root images (10 per class).
\item The repair test retrains B and C from the initialization, partition, and sampling of the main experiment and saves the model at the end of the window. All 30 retrained endpoints are within 5 points of the original ones. Fine-pruning ranks the channels of the last convolutional stage by their mean activation on the root data and prunes until clean accuracy drops by more than one point, reverting the last channel, then fine-tunes for 5 epochs. Fine-tuning alone uses SGD (learning rate 0.01, momentum 0.9) for 5 epochs. Neural Cleanse was adapted from the authors' Keras code to PyTorch with 1{,}000 fixed optimization steps on three root mini-batches per step and no early stopping. The reverse-engineered trigger reached 99\% target success on all 30 models, and unlearning trains 5 epochs on the 100 root images with the trigger applied and the original labels kept.
\item In the FCBA-style test the first four attackers are those of the main experiment and ten more are drawn per seed from the remaining clients. All four arms share one sampling sequence, and all 14 attackers are removed after the window. The 14 nonempty proper subsets are assigned in a fixed order. B14-q2 allocates poisoned samples as B-q4 does.
\item The $E_{\mathrm{adv}}$ test reruns B and C with $E_{\mathrm{adv}}\in\{4,6\}$ and reuses the main-experiment runs for $E_{\mathrm{adv}}=2$.
\end{enumerate}

\paragraph{Failures and interventions} In the original implementation, all 20 CIFAR-10/ResNet-18 runs with FLAME (arms B, C, D, and the clean arm N) stopped after 6 to 8 rounds with non-finite evaluation outputs. They were not rerun under that implementation, and the corrected runs replace them in Table~\ref{tab:emp-defense}. One run without defense (CIFAR-10/ResNet-18, arm D, random, seed 5) was interrupted by a process exit at round 292 and resumed once from its round-275 checkpoint. The replayed rounds reproduced the original evaluation and update logs exactly. One completed run (MNIST/CNN, arm D, FLAME, seed 4) failed only when its completion status was written to disk and was recorded as completed from its existing logs without retraining. Five pretraining checkpoints ended slightly below the accuracy threshold (MNIST/MLP seed 4: 97.96\%; FMNIST/CNN seed 2: 84.97\%; FMNIST/MLP seeds 2 and 3: 84.89\% and 84.92\%; CIFAR-10/VGG-11 seed 4: 73.75\%) and were used as frozen.

\paragraph{Resource audit} Of the 120 paired B/C/D triples, 115 completed and have exactly equal $U$, $P$, and $\Omega$. The other five include a failed FLAME run of the original implementation. All 15 corrected CIFAR-10 FLAME triples completed with equal $U$, $P$, and $\Omega$. Per-run resources, including modified pixel positions and channel values, are in the supplementary data files.

\end{document}